\documentclass{aa}  

\usepackage{graphicx}
\usepackage{url}
\usepackage{txfonts}
\usepackage[normalem]{ulem}

\begin{document}

   \title{Long-term optical and X-ray variability of the Be/X-ray binary H~1145--619: Discovery of an ongoing retrograde density wave}
  \titlerunning{Long-term optical and X-ray variability of the Be/X-ray binary H~1145--619}

   \author{J. Alfonso-Garz\'on
          \inst{1}
          \and J. Fabregat \inst{2} \and P. Reig \inst{3,4} \and J. J. E. Kajava \inst{5,6} \and C. S\'{a}nchez-Fern\'{a}ndez \inst{6} \and 
          L. J. Townsend \inst{7} \and J. M. Mas-Hesse \inst{1} \and S. M. Crawford \inst{8} \and P. Kretschmar \inst{6}  \and M.J. Coe \inst{9}
          }

   \institute{Centro de Astrobiolog\'{\i}a -- Departamento de Astrof\'{\i}sica
              (CSIC-INTA), Camino Bajo del Castillo s/n, E-28692 Villanueva de la Ca\~nada, Spain \\ \email{julia@cab.inta-csic.es}
         \and {Observatorio Astron\'omico, Universidad de Valencia, Catedr\'atico Jos\'e Beltr\'an 2, 46980 Paterna, Spain} \and{IESL, Foundation for Research and Technology-Hellas, 71110, Heraklion, Greece} \and{Physics Department, University of Crete, 71003, Heraklion, Greece} \and{Finnish Centre for Astronomy with ESO (FINCA), University of Turku, V\"{a}is\"{a}l\"{a}ntie 20, FIN-21500 Piikki\"{o}, Finland} \and{European Space Astronomy Centre (ESAC), Camino Bajo del Castillo s/n, E-28692 Villanueva de la Ca\~nada, Spain} \and{Department of Astronomy, University of Cape Town, Private Bag X3, Rondebosch, 7701, South Africa} \and{South African Astronomical Observatory, Observatory Rd, Observatory, Cape Town, South Africa 7935} \and{School of Physics and Astronomy, University of Southampton, Highfield, Southampton SO17 1BJ, UK}\\
             }

   \date{Accepted: 2 September 2017.}

  \abstract
   {Multiwavelength monitoring of Be/X-ray binaries is crucial to understand the mechanisms producing their outbursts. H~1145--619 is one of these systems, which has recently displayed X-ray activity.}
   {We investigate the correlation between the optical emission and  X-ray activity to predict the occurrence of new X-ray outbursts from the inferred state of the circumstellar disc. }
   {We have performed a multiwavelength study of H~1145–619 from 1973 to 2017 and present here a global analysis of its variability over the last 40 years. We used optical spectra from the SAAO, SMARTS, and SALT telescopes and optical photometry from the Optical Monitoring Camera (OMC) onboard \textit{INTEGRAL} and from the All Sky Automated Survey (ASAS). We also used X-ray observations from \textit{INTEGRAL}/JEM--X, and IBIS to generate the light curves and combined them with \textit{Swift}/XRT to extract the X-ray spectra. In addition, we compiled archival observations and measurements from the literature to complement these data.}
   {Comparing the evolution of the optical continuum emission with the H$\alpha$ line variability, we identified three different patterns of optical variability: first, global increases and decreases of the optical brightness, observed from 1982 to 1994 and from 2009 to 2017, which can be explained by the dissipation and replenishment of the circumstellar disc; second, superorbital variations with a period of P$_{superorb}\approx$590~days, observed in 2002--2009, which seems to be related to the circumstellar disc; and third, optical outbursts, observed in 1998--1999 and 2002--2005, which we interpret as mass ejections from the Be star. We discovered the presence of a retrograde one-armed density wave, which appeared in 2016 and is still present in the circumstellar disc.}
   {We carried out the most complete long-term optical study of the Be/X-ray binary H 1145-619 in correlation with its X-ray activity. For the first time, we found the presence of a retrograde density perturbation in the circumstellar disc of a Be/X-ray binary.}


   \maketitle
%

\section{Introduction}\label{sec:int}

Be/X-ray binaries (Be/XBs) consist of a neutron star (NS) orbiting a Be star, which expels matter from the photosphere to create a quasi-Keplerian disc around its equatorial plane. Emission lines and strong infrared excesses are usually observed in these systems owing to the presence of these circumstellar discs. Sometimes, during periastron passages the NS gravitational and magnetic fields channel some of the disc gas into the NS magnetic poles, thereby producing high-energy emission and X-ray outbursts \citep{reig2011}. Be/XBs can exhibit two types of X-ray outbursts: type I or normal outbursts with moderate intensity outbursts (L$_{X}\approx$ 10$^{36}$--10$^{37}$\,erg\,s$^{-1}$) that occur close to the periastron passage of the neutron star; and type II or giant outbursts with (L$_{X} >$~10$^{37}$\,erg\,s$^{-1}$), which can occur at any orbital phase.

The X-ray pulsar H~1145--619 was discovered in 1972 by Uhuru \citep{giacconi1972}. Pulsations from this source, together with pulsations from 1E~1145.1--6141 (17$\arcmin$ away), were detected in 1977 with \textit{Ariel~V} \citep{white1978}. The two sources were resolved by the \textit{Einstein} observatory, identifying H~1145--619 as the source of the 292~s pulsations \citep{white1980, lamb1980}.  A more precise value of the pulse period (292.4~s) was provided by \citet{nagase1989}. The optical counterpart of H~1145--619, Hen~3--715, was discovered in 1978 \citep{dower1978} and classified as a 9-mag B1~Ve \citep{stevens1997}. This system is one of the closest Be/XBs, as it is located at 3.1$\pm$0.5\,kpc from Earth \citep{stevens1997}. It is a long-period eccentric binary (e$>$0.5; \citealt{reig2011}) with an orbital period P$_{\rm orb}$ = 186.68$\pm$0.05\,d \citep{wilson-hodge1999}.

\begin{figure*}[t]
  \begin{center}
    \includegraphics[width=\textwidth]{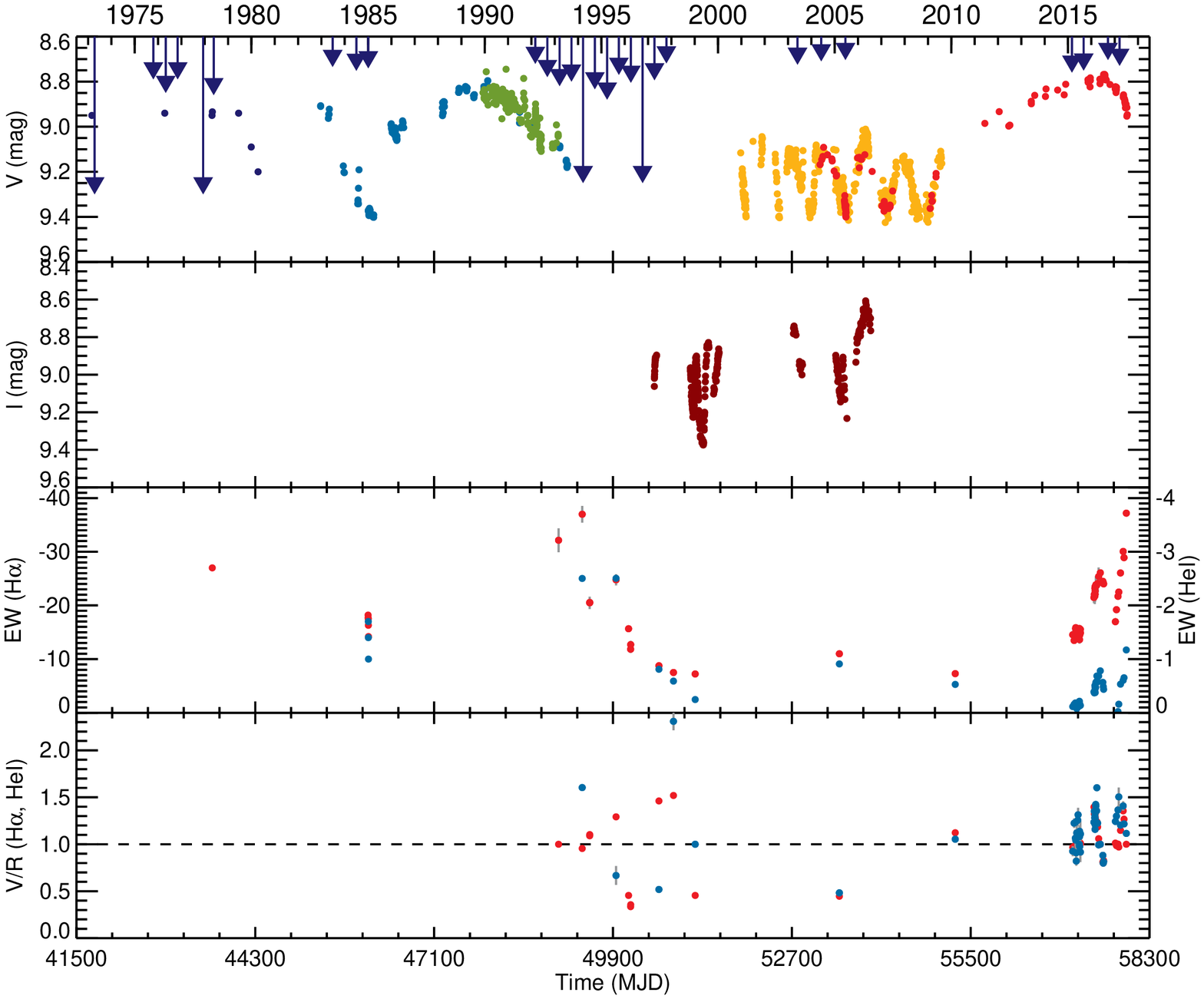}
      \end{center}
\caption[Light curves of H~1145--619]{Optical light curves of H~1145--619 from 1973 to 2015. The X-ray outbursts are indicated with dark blue arrows with lengths proportional to the outburst fluxes. First panel: $V$-- Johnson optical light curves. Dark blue points come from the literature \citep{jones1974,hammerschlag-hensberge1980,pakull1980}, light blue points are from the ESO Long-Term Photometric Variable Project \citep{sterken1995}, green points are from Hipparcos \citep{perryman1997}, orange points come from ASAS--3, and the red points from INTEGRAL/OMC data. Second panel: $I$--band photometric points from ASAS--2 are shown. Third panel:  EW for H$\alpha$ (red points) and He~{\sc i}~6678\,\AA\,(blue points) are shown. We included 1978 and 1985 measurements by \citet{jones1974} and \citet{cook1987-a}, respectively, we re-analysed the observations from 1993 to 1996 published in \citet{stevens1997}, and we present in this work new data and results from 1997 to 2016. Fourth panel: V/R for the same spectra (see text) are shown.}\label{fig:H1145}
\end{figure*}

\citet{priedhorsky1983} analysed the X-ray observations of H~1145-619 from 1969--1976 by the \textit{Vela 5b} satellite, which revealed an outburst of 600\,mCrab (3--12\,keV)  from the source in April 1973. After that, H~1145-619 displayed several X-ray outbursts in 1975--1978, which were observed with \textit{Ariel~V} and \textit{Einstein} \citep{white1980,watson1981}; one of these outbursts reached 600 mCrab in December 1977 \citep{white1980}. X-ray activity was observed again in 1983--1985 with \textit{EXOSAT} \citep{cook1987-a}. After that, nine consecutive outbursts between 1992 and 1997 were observed by \textit{CGRO}/BATSE. Two of these outbursts reached peak fluxes of 550 mCrab (in the 20--40\,keV band) in March 1994 and October 1996 \citep{wilson-hodge1999}. Three of these nine outbursts were also detected by \textit{RXTE}/ASM \citep{corbet1996}. Optical, infrared, and X-ray variability from this epoch were analysed by \citet{stevens1997}, who compiled a 13-yr optical and X-ray history of the system.

From the late 90s to 2015, the source was believed to be in quiescence in X-rays, except for a single detection in 2003 \citep{filippova2005}. In 2015, two X-ray outbursts, coinciding with periastron passages and with intensities of $\sim$50\,mCrab and $\sim$36\,mCrab in the 15--50~keV band, were observed by \textit{MAXI}/GSC and \textit{Swift}/BAT \citep{mihara2015,nakajima2015-a}, suggesting renewed X-ray activity of the source.

In this paper, we aim to understand the correlation between the optical variability and the X-ray activity of the source by performing a comprehensive multiwavelength analysis of the available data of the source.

In Sect.~\ref{sec:obs}, we detail the observations we used in this work. In Sect.~\ref{sec:sptype}, we present the results from the spectral classification and determination of the projected rotational velocity v~$\sin i$. In Sect.~\ref{sec:longt}, we investigate the variability observed with the optical photometric, spectroscopic, and X-ray observations and present the modelling of the X-ray spectrum. In Sect.~\ref{sec:dis} we discuss the results derived in this analysis, and in Sect.~\ref{sec:con}, we summarise the main conclusions we obtained from this work.

\section{Observations} \label{sec:obs}

To analyse the correlation between the optical variability and the X-ray activity of the system, we have made a compilation of the optical and X-ray archival data available for this source since its first known outburst in 1973 to the present (see Fig.~\ref{fig:H1145}). 

\subsection{INTEGRAL}

The INTErnational Gamma-Ray Astrophysics Laboratory (\textit{INTEGRAL}; \citealt{winkler2003}) is a European Space Agency (ESA) mission launched in 2002 in cooperation with the Russian Federal Agency (FKA) and the National Aeronautics and Space Administration (NASA). Data from three \textit{INTEGRAL} instruments were used in this work: the Imager on Board the \textit{INTEGRAL} Satellite (IBIS; \citealt{ubertini2003}), the Joint European X-ray Monitor (JEM-X; \citealt{lund2003}), and the Optical Monitoring Camera (OMC; \citealt{mas-hesse2003}). These instruments provide simultaneous observations of selected sources in hard X-rays (15\,keV--10\,MeV), soft X-rays (3--35\,keV), and in the optical Johnson $V$ band, respectively. We retrieved and analysed archival \textit{INTEGRAL} data of H~1145--619 from 2003 to 2017. All the \textit{INTEGRAL} data were reduced with the Off-line Science Analysis (OSA) software version 10.1.

\subsection{Optical photometry}

We have compiled $V$--band optical measurements from the literature from the 70s--80s \citep{jones1974,hammerschlag-hensberge1980,pakull1980}, and optical photometric light curves from ESO's Long-Term Photometric Variable Project \citep{sterken1995} and from \textit{Hipparcos} \citep{perryman1997}. To convert the Hipparcos H$_{p}$ magnitudes into $V$-- Johnson magnitudes, we used the transformation by \citet{harmanec1998}. We used $V$--Johnson photometric observations from \textit{INTEGRAL}/OMC from 2003 to 2017. In addition, $I$--band and $V$--band light curves from 1997 to 2009 from the All Sky Automated Survey (ASAS; \citealt{pojmanski1997}) have also been used in our analysis. Typical uncertainties are around 0.01\,mag for OMC and in the range 0.03--0.07\,mag for the $V$--band and $I$--band ASAS observations.

\subsection{Optical spectroscopy}

We used in this work spectroscopic measurements from the 1985 outburst by \citet{cook1987-a}. We re-analysed optical spectroscopic observations from \citet{stevens1997} and unpublished observations from 1993--2010. All of these observations were observed with the 1.9 m telescope at the South African Astronomical Observatory (SAAO), except the observation from March 1996, which was observed with the Anglo-Australian Telescope (AAT). The SAAO spectra were taken with the ITS spectrograph and RPCS (Reticon) detector (from 1993 to 1996), and with the SIT 1 and 2 CCD detectors (from 1997 to 2010) using grating No. 5 (1200 lines/mm). The 1200 lines per mm reflection grating blazed at 6800\,\AA\,was used with the SITe CCD, which is effectively 266$\times$1798 pixels in size, yielding a wavelength coverage of 6200\,\AA\, to 6900\,\AA. The pixel scale in this mode was 0.42\,\AA/pixel.

We also performed spectroscopic observations of the system with the high-resolution spectrometer CHIRON on the 1.5 m SMARTS telescope in Cerro Tololo (Chile) in the periods March--July 2015 and February--April 2016. Observations from SMARTS/CHIRON cover the wavelength range of 4150--8800\,\AA\  with a nominal spectral resolution R$=$79,000. We acquired 10 spectra with exposure times of 3$\times$600\,s in 2015 and 15 spectra with exposure times of 450\,s in 2016. The SMARTS/CHIRON data were processed through the SMARTS reduction pipeline \citep{tokovinin2013}.

During the same epoch, we also observed the source with the High Resolution Spectrograph (HRS, \citealt{crause2014}) on the Southern African Large Telescope (SALT) in April--July 2015 and June--July 2016. Observations from SALT/HRS were performed in both low-resolution (LR) and high-resolution (HR) modes. The nominal effective resolving power of each mode is R$=$16,200 and R$=$69,200 respectively. A total of 9 LR mode and 3 HR mode spectra were obtained, with exposure times of 270\,s (LR) and 1200\,s (HR), covering a wavelength range of 3800--8900\,\AA.  Basic data reductions of SALT/HRS observations were completed using the {\it ccdproc} package \citep{craig2015}. The wavelength calibration and extraction were accomplished using the {\it pyhrs} package \citep{crawford2015}.

From January to July 2017, we also observed the source with the Robert Stobie Spectrograph (RSS) on SALT. We used the long-slit mode with the PG2300 grating, covering a spectral range of 6175--6989\,\AA\, with R$=$25,400 at $\lambda$=6500\,\AA. The RSS data were reduced using PySALT \citep{crawford2010}.

\subsection{X-ray spectroscopy}

The \textit{Swift} Gamma-Ray Burst Mission \citep{gehrels2004} is a NASA mission with Italian and UK participations that was launched in 2004. In this work, we used data from one of its instruments, the Swift's X-Ray Telescope (XRT; \citealt{burrows2000}), which provides fluxes, spectra, and light curves in the 0.2--10\,keV band. We analysed the X-ray spectrum of the source, combining data from \textit{INTEGRAL} observations with two \textit{Swift}/XRT spectra of H~1145--619  from March 2015 and September 2016. The spectral data were obtained with the XRT generator tool \citep{evans2009}, and were then binned to have at least 20 counts per bin using the {\sc grppha} tool.

\section{Spectral classification and determination of v~$\sin i$}\label{sec:sptype}

First of all, and in order to be able to subtract the stellar contribution correctly when needed, we performed a spectral classification of the optical companion of the system. For this analysis we selected the 2015 spectra from SALT/HRS. The HRS spectra have a better spectral coverage of the blue region and S/N than Chiron and, as described in Sect.~\ref{sub:opt_spec}, the 2015 observations are less affected by the emission from the circumstellar disc than those from 2016.

We followed the criteria from \citet{walborn1990}. In our spectra, the C~{\sc iii}~4650\,\AA\ line is deeper than the He~{\sc ii}~4686\,\AA\, and Si~{\sc iv}~4089\,\AA\ lines, so the star has to be cooler than B0. Moreover, the He~{\sc ii}~4200 and 4541 lines appear in absorption, which implies that it is warmer than B1. From the relation Si~{\sc iv}~4089/He~{\sc i}~4026, 4144 and from Si~{\sc iii}~4552/He~{\sc i}~4387, the luminosity class should be close to IV. From this criteria, we conclude that the optical companion of H~1145--619 is a B0.2--0.5 IV star, although the intensity of the He~{\sc ii}~4648 line and the relation He~{\sc ii}~4541/Si~{\sc iii}~4552 favour the earlier type.

An alternative method to determine the spectral type is by comparing the observed spectrum with those from model atmospheres. We used the TLUSTY BSTAR2006 models \citep{lanz2007} and explored the parameter space with T$_{eff}$ between 19000 and 30000~K and $\log g$ between 3.25 and 4.0 with steps of 1000~K and 0.25, respectively, assuming solar metallicity. The synthetic spectra were broadened with projected rotational velocities from v~$\sin i$ = 260~km/s to v~$\sin i$ = 320~km/s with a 10~km/s step and we compared them visually with our observations. The best result was obtained for T$_{eff}$ = 29000~K, $\log g$ = 3.25 and v~$\sin i$ = 300~km/s. These physical parameters correspond to a spectral type of B0.2~III, which is in very good agreement with our previous classification but different from the classification given by \citet{stevens1997} of B1~V star.

We used the blue part of the SALT/HRS spectra to determine the v~$\sin i$ independently  with the method described in \citet{steele1999}. We fitted Gaussian profiles to the He~{\sc i} lines at 4026~\AA, 4143~\AA, 4387~\AA, and 4471~\AA. Combining the results from the four lines, we obtained a value of v~$\sin i$ = 280$\pm$30~km/s, which is in agreement with the calculations from the broadened synthetic models.

Assuming for a B0.2\,III star values of M$_{\star} \approx$\,18.5~M$_{\sun}$ and R$_{\star}$ $\approx$ 14~R$_{\sun}$, the critical rotational velocity would be
 \begin{equation}
  V_{break} = \sqrt{\frac{2}{3}\times\frac{GM_{\star}}{R_{\star}}}\approx 400~km/s
\label{eq:vbreak}
  .\end{equation}
  
 We assumed that the radius obtained from the spectral type and luminosity is representative of the polar radius of the star as, due to gravity darkening, for near critical rotators a large percentage of the photospheric light used to derive the radiative parameters comes from the polar regions. With this assumption, the factor 2/3 in eq.~\ref{eq:vbreak} comes from the oblateness of R$_{eq}$ = 2/3 R$_{pole}$ for critical solid body rotation \citep{rivinius2013}. On the other hand, it is now well established that Be stars are fast rotators with a distribution of rotational velocities centred about 80\% of the critical value \citep{chauville2001, rivinius2006, meilland2012}. Assuming that H~1145--619 rotates at 80\% of the critical velocity obtained above and considering the measured value of v$\sin i$ = 300 km/s, we estimated an inclination angle $i \approx$\,70$\degr$.

\begin{figure}[h]
\centering
\includegraphics[width=8.5cm]{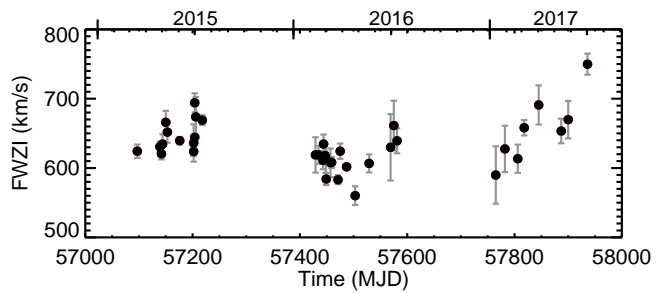}\\
\caption[]{Full width at zero intensity of the He{\sc i}~6678 line.}
\label{fig:dV_He_BW}
\end{figure}

To look for variability on the inclination, we measured the full width at zero intensity (FWZI) for the He{\sc i} line. While for the Balmer lines this quantity is affected by the electron gas temperature due to scattering, for the He{\sc i} line is much more likely to be purely kinematic and should be close to the projected rotational velocity. In Fig.~\ref{fig:dV_He_BW}, these values are plotted from 2015 to 2017. The typical value in 2015--2016 is around 620\,km/s, which implies velocities of 310\,km/s. Since these values remain stable, the inclination should be constant in this period. In 2017, the base width begins to rise smoothly, which could be interpreted as an increase in the inclination of the disc, but more observations are needed to confirm this trend.

\section{Long-term optical and X-ray evolution}\label{sec:longt}

As described in Sect.~\ref{sec:obs}, with the aim of understanding the physical mechanisms producing the observed variability at different wavelengths, we performed a compilation of the available optical and X-ray archival data and measurements in the literature and combined these data with our new observations. An overall picture, showing the optical light curves of H~1145--619, the optical spectroscopic measurements, and all the X-ray outbursts detected from 1973 to 2017 is shown in Fig.~\ref{fig:H1145}. The $V$-- and $I$--band photometric light curves are showed in the first and second panels, respectively. The equivalent widths (EW) and V/R of H$\alpha$ (in red) and He~{\sc i}~6678\,\AA\,(in blue), are shown in the third and fourth panels, respectively (see Table~\ref{tab:spec_obs}). The V/R was calculated as the ratio of the relative intensity at the blue (V) and red (R) emission peaks over the continuum (when double-peaked profiles are present). The X-ray outbursts from the literature and those identified in this work are indicated with dark blue arrows with lengths proportional to the outburst X-ray fluxes (see Table~\ref{tab:xray_outb}).

\begin{figure}
\vspace{0.3cm}
\centering
\includegraphics[width=8.7cm]{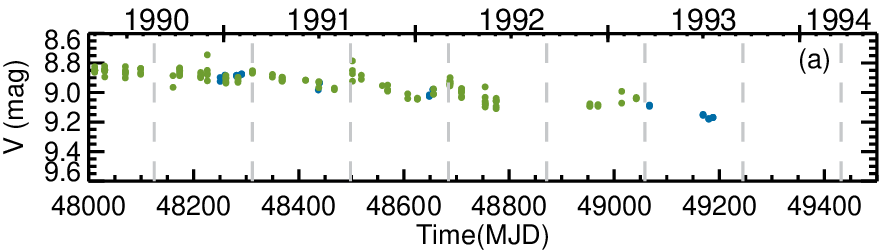}\\
\vspace{0.2cm}
\includegraphics[width=8.7cm]{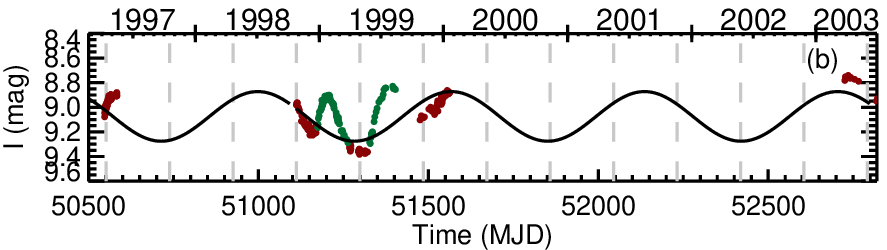}\\
\vspace{0.2cm}
\includegraphics[width=8.7cm]{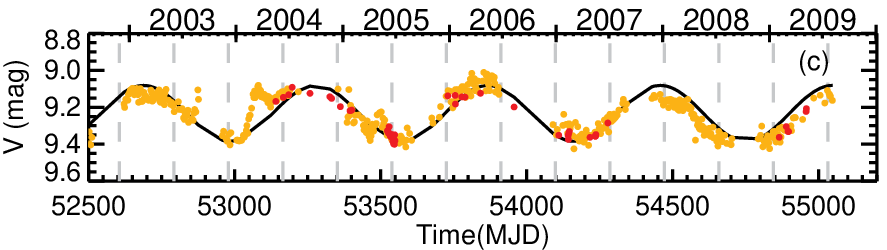}\\
\vspace{0.2cm}
\includegraphics[width=9.1cm]{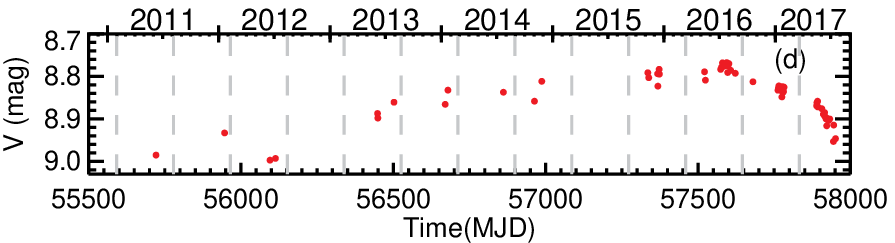}
\caption[]{Optical light curves of H~1145--619 showing different optical behaviour in each period. From top to bottom, (a): $V$--band light curve is represented. Light blue points come from the ESO Long--Term Photometric Variable Project catalogues \citep{sterken1995}, green points are from Hipparcos \citep{perryman1997}. (b): $I$--band data from ASAS--2 are represented; the optical outbursts are indicated in green and observations affected by superorbital variations in brown; see text. (c): $V$--band data from ASAS--3 (orange points) and from INTEGRAL/OMC data (red points) are shown. (d): INTEGRAL/OMC data are shown. Black solid lines in the second and third plots indicate the sinusoidal fit to the superorbital variations. The periastron passages were calculated using the ephemeris P$_{\rm orb}$ = 186.68 d; T$_{0}$ = MJD 48\,871.6 \citep{wilson-hodge1999} and are indicated with grey dashed lines (they are indicated in the subsequent light curves in the paper).}
\label{fig:opt_trends}
\end{figure}

\subsection{Optical photometry}\label{subsec:opt_phot}

The optical light curves of the system present a very intriguing behaviour, showing different types of variability along the 40 years of available observations analysed in this work. A detailed view of each epoch is shown in Fig.~\ref{fig:opt_trends} and described here.

In the 1989--1994 $V$--band light curve (see Fig.~\ref{fig:opt_trends}a), a smooth decrease in the optical brightness is observed. Some faint optical outbursts are superposed onto this decay, two of which are coincident with the periastron passages in August 1991 and March 1992 (MJD 48498.2 and 48684.9).

In the 1996--2000 $I$--band ASAS--2 light curve (see Fig.~\ref{fig:opt_trends}b), the variability is mainly modulated by the orbital period. When we inspect the light curve in detail and compare it with the 2002--2009 $V$--band light curve, there are hints of two different patterns of variability. First, we identified optical outbursts (marked with green points) that are very bright (around 0.5\,mag) and occur at phases close to the apastron and are temporally separated by about an orbital period. Secondly, the superorbital variations, which are dominant in the 2002--2009 period (see below), seem to be present (brown points and black line).

\begin{figure}[h]
\centering
\hspace*{-0.3cm}
\includegraphics[width=0.48\textwidth]{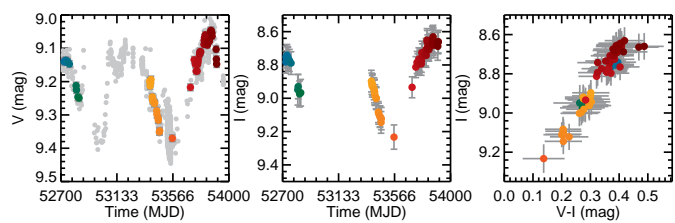}
\caption{$V$--band light curve (left), $I$--band light curve (middle), and colour--magnitude diagram (right) from March 2003 to September 2006. In the $V$--band light curve, all the observations are plotted in grey, but only those simultaneous with the $I$--band observations were used in the colour--magnitude diagram. The points in the outbursts were not used. Different colours were used to represent the temporal scale.}
\label{fig:CMD_superorb}
\end{figure}

From 2002 to 2009, the $V$--band light curve displays optical variations dominated by a sinusoidal superorbital variability of P$_{superorb}\sim$~590\,d (see Fig.~\ref{fig:opt_trends}c). From these epochs, there were simultaneous $V$-- and $I$-- band observations from March 2003 to September 2006. For this period, we represented the evolution of both bands with time and the colour--magnitude diagram (see Fig.~\ref{fig:CMD_superorb}). As can be seen in the colour--magnitude diagram, the system appeared redder when brighter.

Besides these superorbital variations, three optical outbursts in July 2003 (MJD~52850), June 2004 (MJD~53160), and June 2005 (MJD~53500) are detected in the $V$--band light curve. They are similar in shape and duration to the outbursts observed in 1999 in the $I$ light curve, but fainter than those.  These optical outbursts took place in gradually changing orbital phases, starting close to the apastron in 1998--1999 and being closer to the periastron passage in the 2003--2005 events; with the 2005 outburst happening almost at zero phase. Between 2002 and 2006, simultaneous observations in the ASAS--3 $I$ band and ASAS--3 $V$ band were performed (see Fig{\ref{fig:H1145}). The optical outburst in 2005 was observed in both bands, displaying larger amplitude in the $I$ band, with a $V-I$ colour correlated with the flux variations, which is redder when the flux is increasing and bluer when it is decreasing (see Fig.~\ref{fig:CMD}). The points in the colour--magnitude diagram display a loop in which the evolution with time is in clockwise sense.

\begin{figure}[h]
\centering
\hspace*{-0.3cm}
\includegraphics[width=0.48\textwidth]{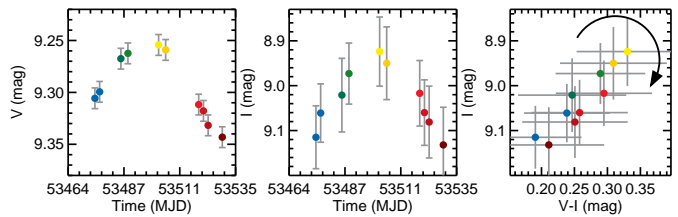}
\caption{$V$--band light curve (left), $I$--band light curve (middle), and colour--magnitude diagram (right) during the 2005 ejection event. Different colours were used to represent the temporal scale.}
\label{fig:CMD}
\end{figure}

From 2010 to mid 2016, the optical flux monotonically increased by $\sim$0.5\,mag, reaching the same optical brightness it showed in the early 90s. From June 2016 to July 2017, the optical flux decreased by $\sim$0.15~mag (see Fig.~\ref{fig:opt_trends}d). This fast decrease is similar to that observed after 1990.

\subsection{Optical spectroscopy}\label{sub:opt_spec}

We used IDL routines to measure the equivalent widths and V/R of the H$\alpha$, H$\beta$, and  He~{\sc i}~6678\,\AA\,lines in all the available optical spectra from 1993 to 2017. These results are summarised in Table~\ref{tab:spec_obs}. The H$\beta$ equivalent widths were calculated to estimate the Balmer decrement, but to make the analysis of the evolution of the emission lines clearer, we just compared the H$\alpha$ and He~{\sc i} lines.

\begin{table*}[!ht]
\caption{Spectroscopic observations of H~1145--619 analysed in this work.}             
\label{tab:spec_obs}      
\centering                          
\small{
\begin{tabular}{l|l|l|rr|rr|rr}        

\hline\hline                 
Date & MJD & Site & EW$_{H\alpha}$ (\AA)& V/R$_{H\alpha}$& EW$_{H\beta}$ (\AA)       & V/R$_{H\beta}$ & EW$_{He{\sc I}}$ (\AA)& V/R$_{He{\sc I}}$ \\    
\hline                        

19930302 & 49048 & SAAO & -32$\pm$2 & 1$^\ast$&     &     &     &    \\
19930304 & 49050 & SAAO &     &     & -4.2$\pm$0.1 & 0.557$\pm$ 0.006 &     &   
 \\
19940307 & 49418 & SAAO & -37$\pm$2 & 0.955$\pm$0.002 &     &     & -2.5$\pm$0.3
 & 1.60$\pm$ 0.03\\
19940702 & 49535 & SAAO & -20$\pm$1 & 1.090$\pm$0.005 &     &     &     &    \\
19940703 & 49536 & SAAO & -20.6$\pm$0.6 & 1.103$\pm$0.007 &     &     &     &   
 \\
19940704 & 49537 & SAAO &     &     & -2.4$\pm$0.2 & 1.29$\pm$ 0.01 &     &    
\\
19950819 & 49948 & SAAO & -25$\pm$1 & 1.292$\pm$0.005 &     &     & -2.5$\pm$0.4
 & 0.7$\pm$ 0.1\\
19960302 & 50144 & AAT & -15.7$\pm$0.5 & 0.455$\pm$0.003 &     &     &     &   
 \\
19960403 & 50176 & SAAO & -11.8$\pm$0.7 & 0.34$\pm$0.02 &     &     &     &    
\\
19960404 & 50177 & SAAO & -12.7$\pm$0.7 & 0.356$\pm$0.005 &     &     &     &   
 \\
19970620 & 50619 & SAAO & -8.8$\pm$0.2 & 1.461$\pm$0.007 &     &     & -0.81
$\pm$0.03 & 0.52$\pm$ 0.01\\
19980203 & 50847 & SAAO & -7.5$\pm$0.2 & 1.519$\pm$0.007 &     &     & -0.59
$\pm$0.02 & 2.31$\pm$ 0.09\\
19990109 & 51187 & SAAO & -7.2$\pm$0.2 & 0.455$\pm$0.005 &     &     & -0.25
$\pm$0.01 & 1$\pm$ 0\\
20050315 & 53444 & SAAO & -11.0$\pm$0.2 & 0.446$\pm$0.002 &     &     & -0.91
$\pm$0.02 & 0.48$\pm$ 0.01\\
20060508 & 53863 & SAAO &     &     & -0.47$\pm$0.02 & 0.82$\pm$ 0.02 &     &   
 \\
20100303 & 55258 & SAAO & -7.3$\pm$0.2 & 1.123$\pm$0.009 &     &     & -0.53
$\pm$0.04 & 1.05$\pm$ 0.03\\
20150316 & 57097 & SMARTS/Chiron & -14.5$\pm$0.7 & 0.967$\pm$0.005 & -1.4$\pm$
0.1 & 1.003$\pm$ 0.008 & -0.12$\pm$0.04 & 0.93$\pm$ 0.05\\
20150407 & 57119 & SMARTS/Chiron & -13.5$\pm$0.3 & 0.932$\pm$0.002 & -1.2$\pm$
0.1 & 0.89$\pm$ 0.02 & -0.16$\pm$0.04 & 1.22$\pm$ 0.05\\
20150427 & 57139 & SMARTS/Chiron & -14.4$\pm$0.5 & 0.937$\pm$0.006 & -1.42$\pm$
0.08 & 0.94$\pm$ 0.01 & -0.18$\pm$0.09 & 1.07$\pm$ 0.09\\
20150430 & 57142 & SALT/HRS & -15.7$\pm$0.4 & 0.993$\pm$0.003 & -1.538$\pm$0.004
 & 0.985$\pm$ 0.004 & 0.07$\pm$0.01 & 1.23$\pm$ 0.03\\
20150502 & 57144 & SALT/HRS & -15.9$\pm$0.3 & 0.985$\pm$0.003 & -1.510$\pm$0.002
 & 0.976$\pm$ 0.004 & 0.08$\pm$0.02 & 1.0$\pm$ 0.1\\
20150508 & 57150 & SALT/HRS & -15.3$\pm$0.3 & 0.968$\pm$0.004 & -1.544$\pm$0.004
 & 1.004$\pm$ 0.004 & 0.07$\pm$0.02 & 0.9$\pm$ 0.1\\
20150511 & 57153 & SALT/HRS & -15.6$\pm$0.5 & 0.971$\pm$0.003 & -1.522$\pm$0.003
 & 0.99$\pm$ 0.01 & 0.09$\pm$0.01 & 0.82$\pm$ 0.05\\
20150518 & 57160 & SMARTS/Chiron & -14.7$\pm$0.3 & 0.947$\pm$0.006 & -1.32$\pm$
0.07 & 0.99$\pm$ 0.01 & -0.08$\pm$0.07 & 1.1$\pm$ 0.1\\
20150603 & 57176 & SMARTS/Chiron & -15.2$\pm$0.5 & 0.97$\pm$0.01 & -1.50$\pm$
0.07 & 1.00$\pm$ 0.02 & -0.10$\pm$0.02 & 1.26$\pm$ 0.06\\
20150608 & 57181 & SMARTS/Chiron & -15.1$\pm$0.4 & 0.962$\pm$0.005 & -1.54$\pm$
0.05 & 0.97$\pm$ 0.01 & -0.13$\pm$0.02 & 1.31$\pm$ 0.07\\
20150629 & 57202 & SMARTS/Chiron & -14.8$\pm$0.5 & 0.998$\pm$0.006 & -1.465$\pm$
0.002 & 1.076$\pm$ 0.004 & 0.10$\pm$0.03 & 1.0$\pm$ 0.1\\
20150629 & 57202 & SALT/HRS & -15.7$\pm$0.5 & 1.018$\pm$0.004 & -1.6$\pm$0.1 & 
1.089$\pm$ 0.006 & -0.21$\pm$0.06 & 0.98$\pm$ 0.1\\
20150630 & 57203 & SMARTS/Chiron & -14.7$\pm$0.6 & 0.987$\pm$0.009 & -1.48$\pm$
0.06 & 1.08$\pm$ 0.01 & -0.13$\pm$0.04 & 1.12$\pm$ 0.04\\
20150701 & 57204 & SMARTS/Chiron & -13.6$\pm$0.7 & 0.990$\pm$0.006 & -1.22$\pm$
0.08 & 1.15$\pm$ 0.01 & 0.06$\pm$0.02 & 1.14$\pm$ 0.06\\
20150701 & 57204 & SALT/HRS & -14.6$\pm$0.2 & 1.019$\pm$0.004 & -1.662$\pm$0.003
 & 1.074$\pm$ 0.004 & -0.18$\pm$0.03 & 1.08$\pm$ 0.06\\
20150703 & 57206 & SALT/HRS & -14.5$\pm$0.5 & 1.008$\pm$0.004 & -1.644$\pm$0.001
 & 1.04$\pm$ 0.01 & 0.06$\pm$0.03 & 0.97$\pm$ 0.06\\
20150714 & 57217 & SMARTS/Chiron & -15.5$\pm$0.8 & 1.012$\pm$0.006 & -1.65$\pm$
0.05 & 1.04$\pm$ 0.02 & -0.14$\pm$0.08 & 0.9$\pm$ 0.1\\
20150715 & 57218 & SALT/HRS & -14.9$\pm$0.4 & 1.004$\pm$0.005 &     &     & 0.08
$\pm$0.05 & 1.1$\pm$ 0.1\\
20160211 & 57429 & SMARTS/Chiron & -21$\pm$1 & 1.396$\pm$0.009 & -3.0$\pm$0.2 & 
1.79$\pm$ 0.05 & -0.4$\pm$0.1 & 1.23$\pm$ 0.09\\
20160217 & 57435 & SMARTS/Chiron & -22.0$\pm$0.9 & 1.32$\pm$0.01 & -3.13$\pm$
0.07 & 1.63$\pm$ 0.04 & -0.37$\pm$0.05 & 1.35$\pm$ 0.03\\
20160225 & 57443 & SMARTS/Chiron & -22$\pm$1 & 1.29$\pm$0.01 & -3.26$\pm$0.09 & 
1.80$\pm$ 0.03 & -0.41$\pm$0.09 & 1.28$\pm$ 0.03\\
20160226 & 57444 & SMARTS/Chiron & -23.2$\pm$0.8 & 1.278$\pm$0.007 & -3.3$\pm$
0.1 & 1.82$\pm$ 0.02 & -0.51$\pm$0.05 & 1.30$\pm$ 0.09\\
20160229 & 57447 & SMARTS/Chiron & -23.5$\pm$0.6 & 1.33$\pm$0.01 & -3.35$\pm$
0.09 & 1.89$\pm$ 0.02 & -0.49$\pm$0.04 & 1.16$\pm$ 0.02\\
20160301 & 57448 & SMARTS/Chiron & -22$\pm$1 & 1.318$\pm$0.008 & -3.5$\pm$0.1 & 
1.76$\pm$ 0.02 & -0.46$\pm$0.05 & 1.28$\pm$ 0.03\\
20160302 & 57449 & SMARTS/Chiron & -23$\pm$1 & 1.290$\pm$0.005 & -3.4$\pm$0.1 & 
1.74$\pm$ 0.02 & -0.38$\pm$0.03 & 1.21$\pm$ 0.03\\
20160310 & 57457 & SMARTS/Chiron & -23.6$\pm$0.6 & 1.233$\pm$0.001 & -3.7$\pm$
0.1 & 1.52$\pm$ 0.01 & -0.47$\pm$0.04 & 1.42$\pm$ 0.02\\
20160312 & 57459 & SMARTS/Chiron & -24$\pm$2 & 1.19$\pm$0.01 & -3.8$\pm$0.1 & 
1.45$\pm$ 0.01 & -0.57$\pm$0.07 & 1.41$\pm$ 0.02\\
20160324 & 57471 & SMARTS/Chiron & -24$\pm$1 & 1.195$\pm$0.004 & -3.8$\pm$0.1 & 
1.35$\pm$ 0.01 & -0.58$\pm$0.02 & 1.36$\pm$ 0.03\\
20160328 & 57475 & SMARTS/Chiron & -23.9$\pm$0.7 & 1.240$\pm$0.001 & -4.1$\pm$
0.1 & 1.40$\pm$ 0.01 & -0.69$\pm$0.05 & 1.60$\pm$ 0.04\\
20160409 & 57487 & SMARTS/Chiron & -24$\pm$2 & 1.18$\pm$0.01 & -4.2$\pm$0.1 & 
1.33$\pm$ 0.02 & -0.58$\pm$0.09 & 1.22$\pm$ 0.02\\
20160425 & 57503 & SMARTS/Chiron & -25$\pm$2 & 1.059$\pm$0.006 & -4.6$\pm$0.2 & 
1.17$\pm$ 0.02 & -0.68$\pm$0.05 & 0.99$\pm$ 0.02\\
20160521 & 57529 & SMARTS/Chiron & -26.1$\pm$0.5 & 1$^\ast$& -5.0$\pm$0.2 & 
1.001$\pm$ 0.002 & -0.78$\pm$0.05 & 1$^\ast$ \\
20160630 & 57569 & SALT/HRS & -24.5$\pm$0.6 & 0.810$\pm$0.003 & -4.92$\pm$0.01
 & 0.639$\pm$ 0.006 & -0.57$\pm$0.02 & 0.88$\pm$ 0.02\\
20160706 & 57575 & SALT/HRS & -24.2$\pm$0.6 & 0.800$\pm$0.003 & -4.764$\pm$0.007
 & 0.639$\pm$ 0.005 & -0.49$\pm$0.04 & 0.80$\pm$ 0.02\\
20160712 & 57581 & SALT/HRS & -24.0$\pm$0.5 & 0.826$\pm$0.002 & -4.71$\pm$0.01
 & 0.6681$\pm$ 0.0008 & -0.44$\pm$0.02 & 0.81$\pm$ 0.02\\
20170112 & 57765 & SALT/RSS & -17.0$\pm$0.1 & 1.013$\pm$0.007 &     &     & 0.36
$\pm$0.01 & 1.19$\pm$ 0.04\\
20170129 & 57782 & SALT/RSS & -19.20$\pm$0.08 & 0.985$\pm$0.004 &     &     & 
0.262$\pm$0.003 & 1.33$\pm$ 0.05\\
20170222 & 57806 & SALT/RSS & -21.7$\pm$0.1 & 1.002$\pm$0.005 &     &     & 0.00
$\pm$0.01 & 1.63$\pm$ 0.05\\
20170306 & 57818 & SALT/RSS & -22.49$\pm$0.05 & 0.970$\pm$0.006 &     &     & 
-0.145$\pm$0.008 & 1.32$\pm$ 0.03\\
20170402 & 57845 & SALT/RSS & -26.05$\pm$0.03 & 1.148$\pm$0.006 &     &     & 
-0.540$\pm$0.006 & 1.21$\pm$ 0.01\\
20170514 & 57887 & SALT/RSS & -30.1$\pm$0.4 & 1.36$\pm$0.02 &     &     & -0.621
$\pm$0.004 & 1.412$\pm$ 0.009\\
20170527 & 57900 & SALT/RSS & -28.9$\pm$0.1 & 1.267$\pm$0.003 &     &     & 
-0.657$\pm$0.007 & 1.23$\pm$ 0.02\\
20170702 & 57936 & SALT/RSS & -37.2$\pm$0.3 & 1$^\ast$&     &     & -1.20$\pm$
0.01 & 1.13$\pm$ 0.01\\

\hline        
\end{tabular}
}
\tablefoot{$^{\ast}$ The line shows a single-peaked profile.}
\end{table*}

\subsubsection{Evolution of line profiles }
The line profiles of the H$\alpha$, and He~{\sc i}~6678\,\AA\,lines from 1993 to 2010 are shown in Fig.~\ref{fig:spec_old}.
It can be seen that at the beginning of the X-ray active phase in the 90s (from March 1993 to July 1994), the H$\alpha$ emission was larger and the profiles were single-peaked, while He~{\sc i} was hardly detected in emission (the S/N is only good enough in the spectrum from March 1994). After the bright outburst in 21 March 1994, the H$\alpha$ EW decreased abruptly from EW$\simeq$-37\AA (March 1994) to EW$\simeq$-20\,\AA\, (July 1994). Another bright outburst took place in October 1996. In this case the H$\alpha$ EW decreased from EW$\simeq$-13\,\AA\,(April 1996) to EW$\simeq$-9\,\AA\,(June 1997). In some spectra from this epoch (19 August 1995 and 20 June 1997), the V/R variations are in anti-phase for H$\alpha$ and He~{\sc i}, pointing to fast changes in the structure of the circumstellar disc. In the spectra between both dates, the He~{\sc i} is not detected in emission. From 2000 to 2009, only two spectra were obtained. In March 2015 the H$\alpha$  and and He{\sc i} lines presented double-peaked profiles with V/R=0.45 and 0.48, respectively, and in May 2006 H$\beta$ also presented a double-peaked profile but with V/R=0.82. Another spectra was taken in March 2010, in which the H$\alpha$ and He{\sc i} lines displayed almost symmetric double-peaked profiles.

\begin{figure}[t]
\begin{center}
  \includegraphics[width=3.4cm]{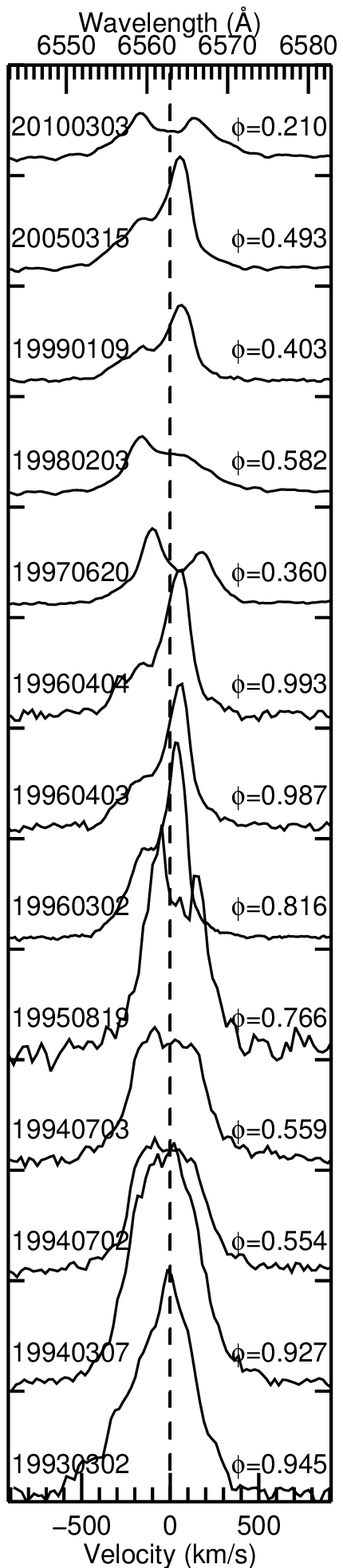}
\includegraphics[width=3.4cm]{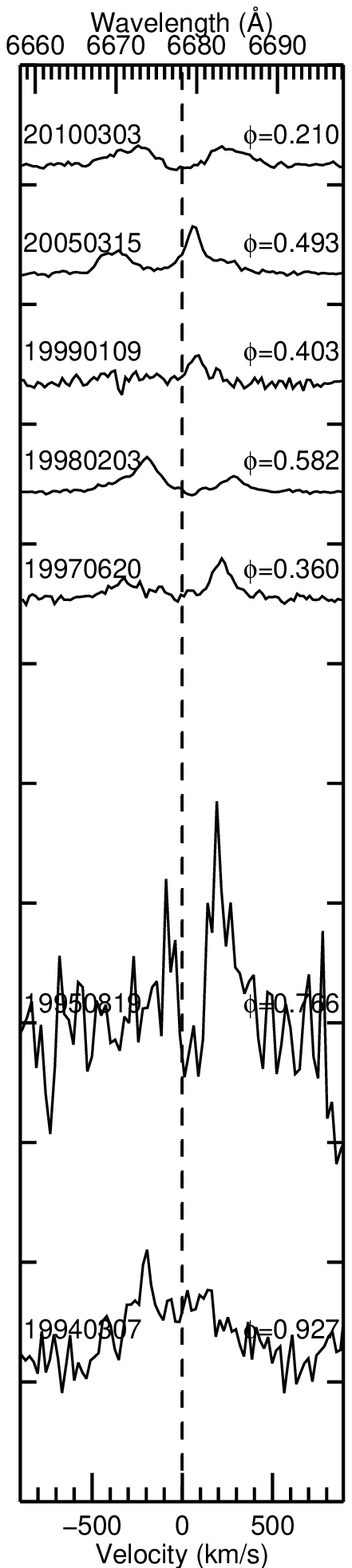}
\caption{H$\alpha$ (left) and He~{\sc i}~6678\,\AA \,(right) line profiles from the optical spectra from the SAAO/1.9\,m telescope of H~1145--619 from March 1993 (bottom) to March 2010 (top). The rest wavelengths for each line are indicated with black dashed lines and the dates and orbital phases are indicated on the sides of each line. The flux scale is arbitrary and different for each panel.}\label{fig:spec_old}
 \end{center}
\end{figure}

The line profiles from the 2015--2017 spectra are shown in Fig.~\ref{fig:wden}. The evolution of the EW and V/R values for both lines is represented in Fig.~\ref{fig:spec_res} (H$\alpha$ in red and He{\sc i} in blue). During the 2015 observations, the EWs did not vary substantially and the V/R ratios stayed close to 1 with double-peaked symmetric profiles of the lines. However, from February 2016 to July 2017, clear V/R variations and EW evolution are found, being more pronounced in 2017. From February to May 2016, the equivalent widths increased significantly, reaching values around $-25$\,\AA\, for H$\alpha$ and $-0.7$\,\AA\,for the He~{\sc i}~6678\,\AA\, line in May 2016. After that, the observed emission from both lines started to decrease. From January to July 2017, the EW increased faster than in 2016, displaying values in the last spectra similar to those measured in the 90s with H$\alpha$ equivalent widths around $-37$\,\AA. In 2016, the V/R values changed from V/R$>$1 (blue peak brighter than the red peak) in February--April 2016 to V/R$<$1 (red peak brighter than the blue  peak) in June--July 2016, passing by a single-peaked profile (in which we adopted V/R$=$1) in May 2016. From January 2017 to July 2017, the line profiles evolved from double-peaked with V/R$\sim$1 (January--March) to double-peaked with V/R$>$1 (April--June), finally displaying single-peaked profile for H$\alpha$ and almost single-peaked profile for the He~{\sc i} line (July 2017).

\begin{figure}[h]
\begin{center}
 \includegraphics[width=9cm]{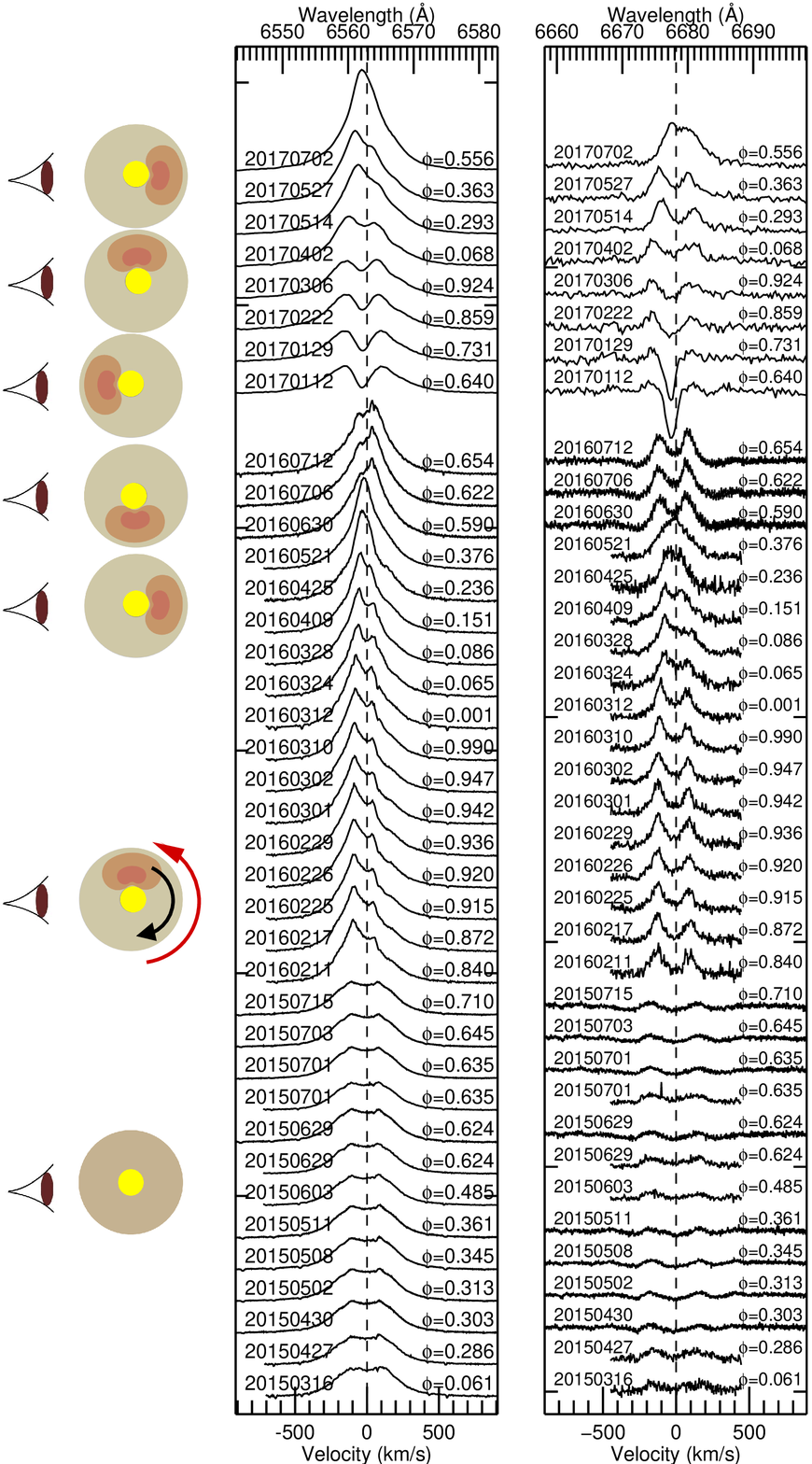}
\caption{Optical high-resolution SMARTS/CHIRON, SALT/HRS, and SALT/RSS spectra of H~1145--619 from March 2015 (bottom) to July 2017 (top). Left panel: H$\alpha$ profiles are shown. Right panel: He~{\sc i}~6678\,\AA \,profiles are shown. Same coding as in Fig.~\ref{fig:spec_old}. The rest wavelengths for each line are indicated with black dashed lines and the dates and the orbital phases are indicated on the sides of each line. The flux scale is arbitrary and different for each panel. The one-armed density scheme has been adapted from \citet{telting1994} and is added on the left part of the figure. The red arrow indicates the sense of the motion of the material in the circumstellar disc and the black arrow indicates the sense of the propagation of the density wave.}
\label{fig:wden}
  \end{center}
  \end{figure}

\begin{figure}[h]
  \begin{center}
  \includegraphics[width=0.48\textwidth]{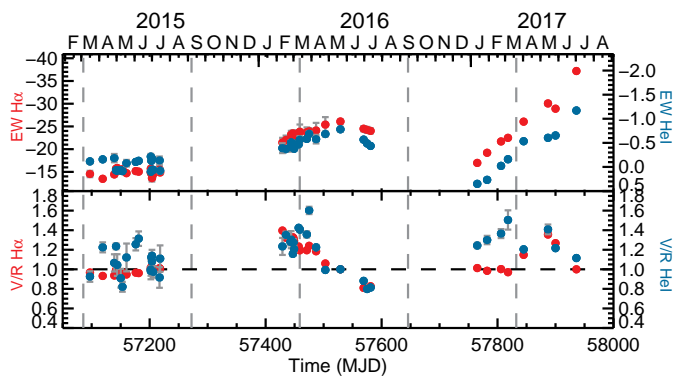}
\caption[EW H1145--619]{Spectroscopic measurements in the 2015--2017 spectra of H~1145--619. Equivalent widths (top) and V/R ratios (bottom) measured for the H$\alpha$ (red filled circles) and He~{\sc i}~6678 (blue filled circles) lines. In most of the cases, the error bars cannot be distinguished because they are smaller than the symbols.}\label{fig:spec_res}
  \end{center}
\end{figure}

\begin{figure}
  \begin{center}
\includegraphics[width=0.48\textwidth]{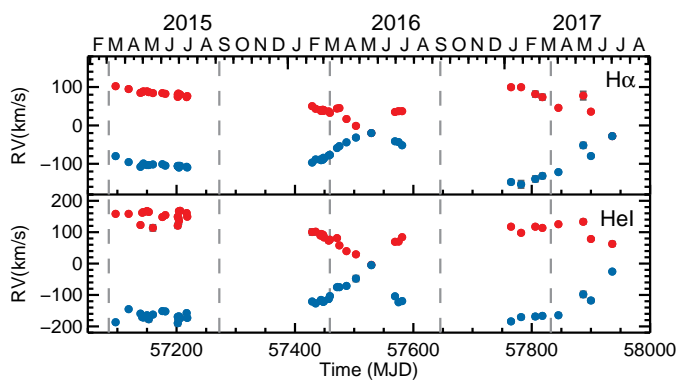}
\caption[EW H1145--619]{Radial velocities from the V (blue filled circles) and R (red filled circles) peaks measured in the H$\alpha$ (top) and He~{\sc i}~6678 (bottom) lines in the 2015--2017 spectra.}\label{fig:spec_RVs}
  \end{center}
\end{figure}

We also measured the radial velocities derived from the wavelengths of the blue and red peaks of each emission line, $RV=(\lambda-\lambda_{0})/\lambda_{0}\times c$, where c is the speed of light and $\lambda_{0}$ is the rest wavelength. The evolution of these RVs is shown in Fig.~\ref{fig:spec_RVs}. The amplitude of the observed velocities is larger for the He~{\sc i} than for H$\alpha$. For both lines, the RVs of the peaks remained almost stable in 2015 with $\Delta$V$\sim$190~km/s for H$\alpha$ and $\Delta$V$\sim$320~km/s for He~{\sc i}. In 2016, the velocities of both peaks displayed symmetrical variability. The RVs of both peaks evolved very fast from $\Delta$V$\sim$150~km/s for H$\alpha$ and $\Delta$V$\sim$225~km/s for He~{\sc i} in February 2016, where the blue peak is brighter than the red peak, until May 2016, when the two peaks could not be distinguished (single-peaked profile). In this case, and taking the observed trend into account, the same velocity (RV$\simeq$0 km/s) was considered for both V and R peaks. From May 2016 to July 2016, $\Delta$V started to increase again; in this case, the red peak was brighter than the blue peak. In 2017, the same behaviour is observed, but with higher amplitude of the variability, evolving from symmetric double-peaked profiles with $\Delta$V$\sim$250~km/s for H$\alpha$ and $\Delta$V$\sim$300~km/s for the He~{\sc i} line; these profiles evolved from less separated blue-dominated double-peaked profiles, to a single-peaked profile for H$\alpha,$ and a still blue-dominated double-peaked profile with $\Delta$V$\sim$100~km/s for the He~{\sc i} in July 2017. As shown in Sect.~\ref{sec:dis}, some of these measurements can be affected by the shell absorption by the emitting material.

As mentioned above, there are two epochs in which the blue peak is brighter than the red peak (February--April 2016 and April--May 2017) (see Fig.~\ref{fig:wden}). There are also two periods in which the lines displayed single-peaked profiles (May 2016 and July 2017). Comparing these epochs and considering that the radial velocities are dominated by the presence of a density wave (see Fig.~\ref{fig:spec_RVs}), we can identify a quasi-periodic motion of the enhanced emission with a period of around P$_{dens-wave}\sim$~420\,d. In Sect.~\ref{sec:dis}, we go further into the interpretation of these results.

\subsubsection{Balmer decrement}\label{ssub:BD}

The strongest emission lines in Be stars are those of the Balmer series of hydrogen. The intensity ratios of the different lines are called Balmer decrements. We calculated the D$_{34}$ Balmer decrement, which is defined as

\begin{equation}
 D_{34}=\frac{I(H\alpha)}{I(H\beta)}=\frac{F_{cont}(H\alpha)\times EW'(H\alpha)}{F_{cont}(H\beta)\times EW'(H\beta)}
,\end{equation}

\noindent where $F_{cont}(H\alpha)$ and $F_{cont}(H\beta)$ are the fluxes at continuum level at the corresponding rest wavelengths and EW'(H$\alpha$) and EW'(H$\beta$) are the equivalent widths of both lines, corrected from the stellar contribution. Both continuum fluxes and stellar contributions to the equivalent widths were measured in the synthetic model described in Sect.~\ref{sec:sptype}.

The obtained values for the D$_{34}$ Balmer decrement range between 1.2--1.5. These values are consistent with the mean Balmer decrement for early Be stars with typical values in the range 1.2--3.2 \citep{dachs1990}. Our values are at the flatter limit of this interval. As a flatter Balmer decrement indicates a denser disc, the circumstellar disc of H~1145--619 should be denser than the average for early Be stars. This is consistent with the by now well-known fact that discs in Be/X-ray binaries are on average denser than the discs of isolated Be stars \citep{reig2016}.

\begin{figure}[h]
\centering
\includegraphics[width=0.48\textwidth]{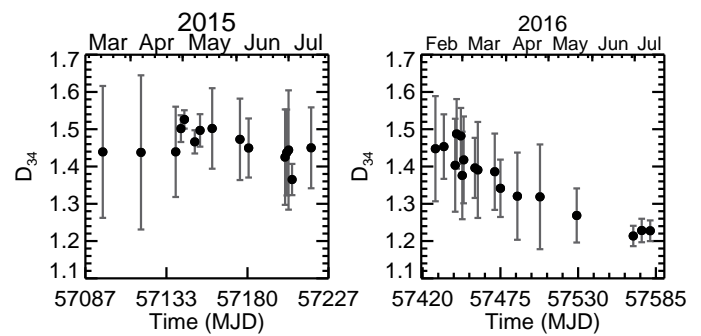}
\caption{Evolution of the D$_{34}$ Balmer decrement during 2015 (left) and 2016 (right).}
\label{fig:Bdec}
\end{figure}

In Fig.~\ref{fig:Bdec} we represented the evolution of the D$_{34}$ Balmer decrement in 2015 and 2016. During 2015 D$_{34}$ remains nearly constant, with some irregular, low amplitude variation around a mean value of 1.45. In 2016, however, there is a clear trend towards a flattening of the decrement from 1.45 to 1.2, indicating a progressive increase of the disc electron density. This could be related with the development and progress of the  density wave in the disc.

We also measured the D$_{34}$ Balmer decrement for the two spectra in the 90s for which we had observations of H$\alpha$ and H$\beta$ in close dates. These spectra were acquired in 2 and 4 March 1993, and in 3 and 4 July 1994 (see Table~\ref{tab:spec_obs}). From these data, we derive D$_{34}=$~1.7 and D$_{34}=$~1.5, respectively, suggesting that the disc was less dense in that epoch than in 2015--2016.

\subsection{X-ray activity}\label{sub:xact}

As we described in Sect~\ref{sec:int}, H~1145--619 was very active in the 70s, 80s, and 90s, displaying several faint and bright X-ray outbursts. The last recorded X-ray outburst from this epoch was detected in September 1996 by \textit{RXTE}/ASM \citep{corbet1996} and \textit{CGRO}/BATSE  \citep{wilson-hodge1999}. After that the source entered into quiescence. 

\begin{figure}[h]
\centering
\hspace{-0.cm}\includegraphics[width=4.4cm]{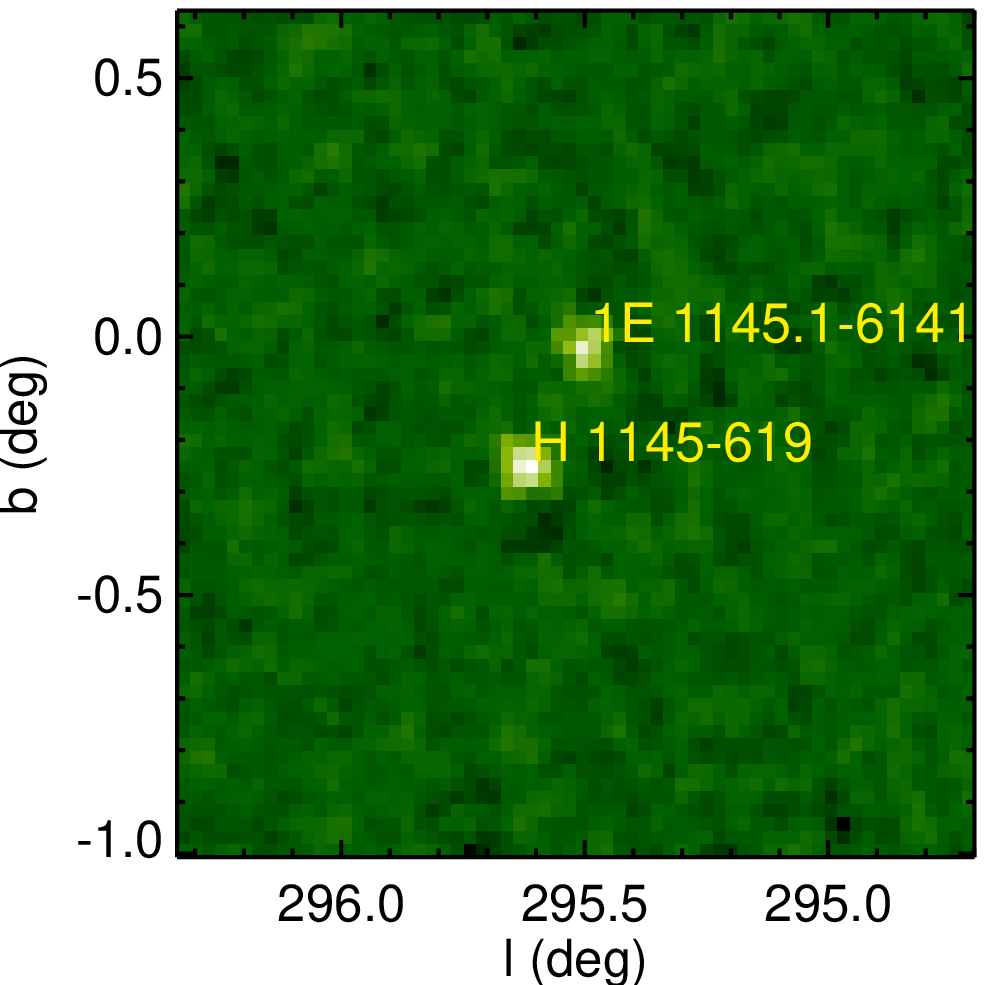}
\hspace{-0.1cm}\includegraphics[width=4.4cm]{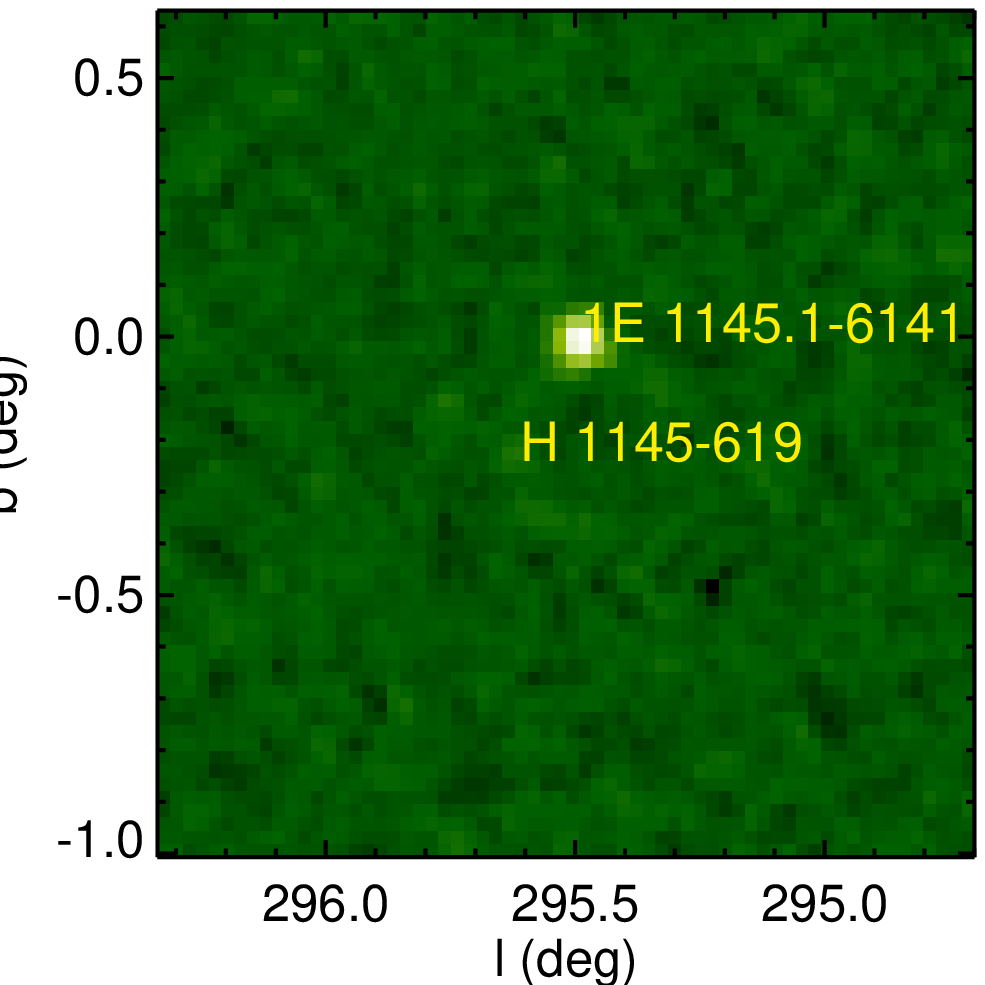}\\
\hspace{-0.cm}\includegraphics[width=4.4cm]{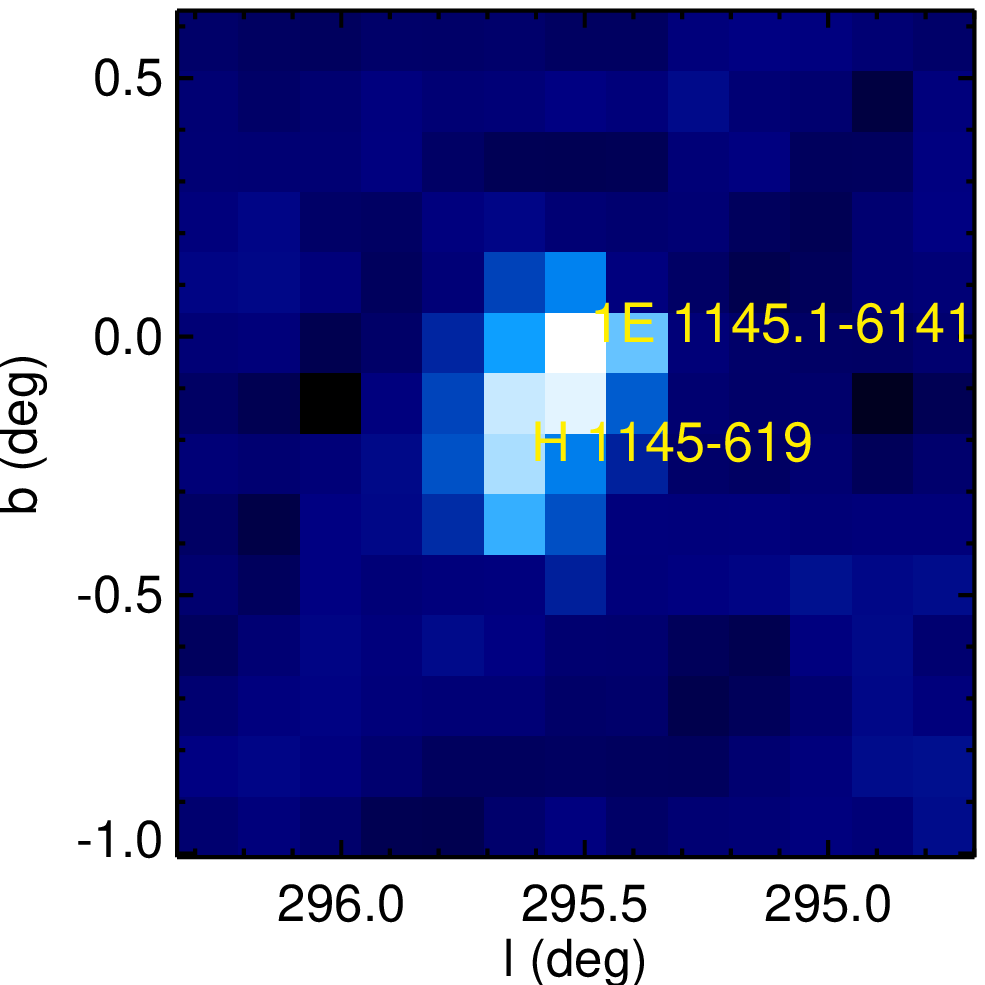}
\hspace{-0.1cm}\includegraphics[width=4.4cm]{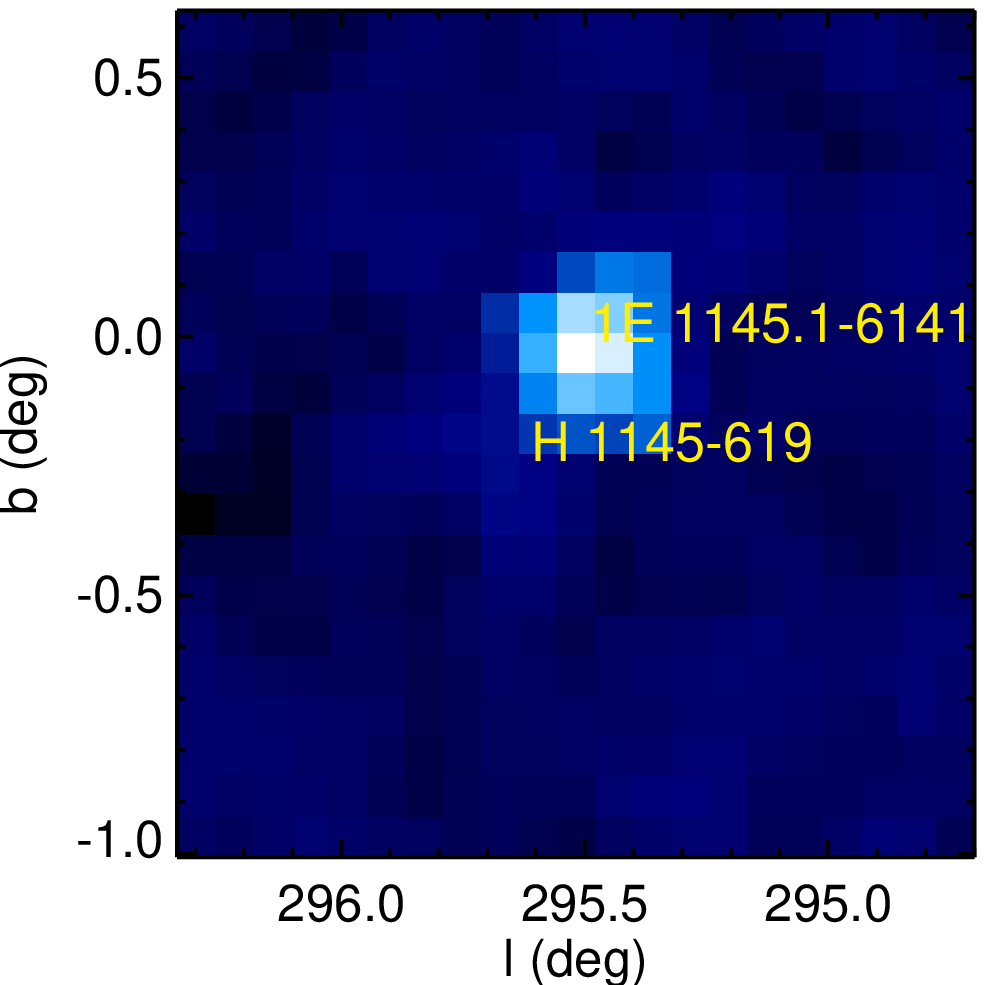}\\
\includegraphics[width=9cm]{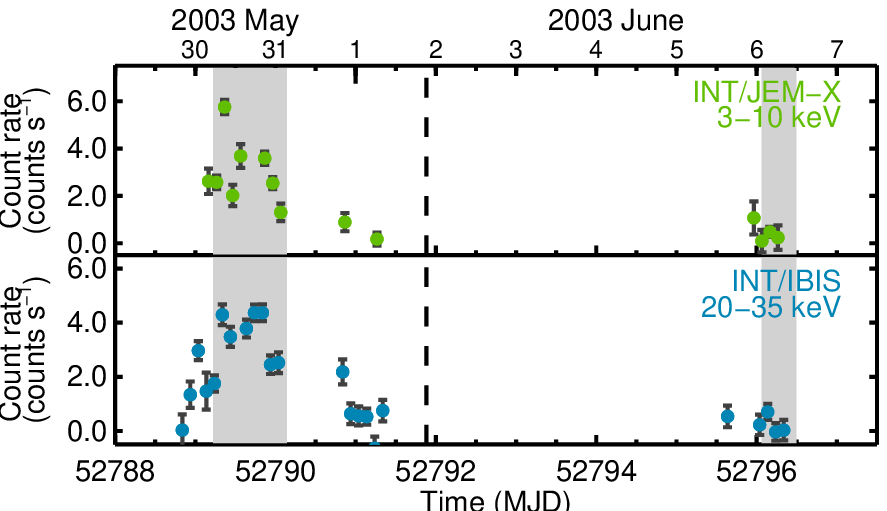}
\caption[JEM-X and IBIS images H~1145--619 in the outburst of 2003]{In green, the JEM-X images in the 3--10 keV energy band and in blue the IBIS/ISGRI images in the 20--35 keV energy range of H~1145--619 in outburst (left) and quiescence (right) in the epoch of the 2003 outburst. The field of view is of 2$\degr\times$2$\degr$ and H~1145--619 is the source located more to the south and east appearing only in the images from the outburst, and the other source appearing in all the images is 1E~1145.1--6141. At the bottom, the corresponding light curves of H~1145--619 are plotted. The data used to extract the images are shown with a grey shadow.}
\label{fig:H1145_imout}
\end{figure}

We have analysed the available \textit {INTEGRAL}/JEM-X and IBIS observations, identifying three outbursts in May 2003, June 2004, and June 2005. The 2003 outburst had been previously analysed by \citet{filippova2005}, while the other two outbursts had not been previously reported. The fluxes of these outbursts were 33, 15, and 7\,mCrab in the 22--40\,keV band (measured with \textit{INTEGRAL}/IBIS), respectively. The intensity of these three outbursts decreased with time, while the duration increased from 2003 to 2005 ($\sim$3 days in 2003, $\sim$10 days in 2004, and hard to measure in 2005). The IBIS angular resolution is 12~arcmin FWHM, hence unless a careful analysis is made; a risk of source confusion with the nearby source 1E~1145.1--6141, 17\arcmin away, cannot be ruled out. We verified the identification of H~1145--619 as the active source at these epochs by generating the IBIS and JEM-X images of the field before and during the outbursts; we also separately extracted the light curves of both sources. In Fig.~\ref{fig:H1145_imout}, images of the 2003 outburst during the epoch in outburst (left) and quiescence (right) are shown. From top to bottom the energy bands are 3--10 keV from JEM-X and 20--35 keV from IBIS/ISGRI.  The brightening of the source in the latter can be observed by comparing the images in quiescence with the images in outburst. In the JEM-X image both sources are resolved and in the images corresponding to the outburst, two point sources are observed. However, in the IBIS/ISGRI images of the outburst epoch, the presence of H~1145--619 is evident, but it is blended with 1E~1145.1--6141 due to the poorer angular resolution. The light curves of H~1145--619 with both instruments are also provided in the figure. The observations used to extract the images are indicated in grey in the respective light curves.

After 2005 and until 2015, no X-ray outbursts were identified. Since then, the source has displayed faint X-ray activity at every almost periastron passage (no X-ray emission was detected during the periastron passage in March 2016). In March and September 2015, two X-ray outbursts in coincidence with the periastron passages and intensities of $\sim$50~mCrab and $\sim$36~mCrab in the 15--50~keV band were observed by \textit{MAXI}/GSC and \textit{Swift}/BAT \citep{mihara2015,nakajima2015-a}, indicating that the source was active again. In 18 September 2016, we observed H~1145--619 with \textit{Swift}/XRT and identified another outburst, which was much fainter than the 2015 outbursts ($\sim$4~mCrab in the 2--10 keV band). In 19 March 2017, the source underwent another normal outburst that was observed by \textit{Swift}/BAT with a flux of $\sim$30~mCrab in the 15--50 keV band. This outburst was also detected with \textit{MAXI}/GSC \citep{nakajima2017}.

A compilation of all the X-ray outbursts detected from this source is provided in Table~\ref{tab:xray_outb}.

\begin{table}[h]
\caption{X-ray outbursts history of H~1145-619.}             
\label{tab:xray_outb}      
\begin{center}
 
\begin{tabular}{c c c r r}        
\hline\hline                 
Date  &   T$_{peak}$(MJD)  & Flux (mCrab)& E band & Ref \\
\hline                        
1973-04-13  & 41785  &  600      &  2--10\,keV &     1     \\
1975-10-19  & 42704  &  90       &  2--10\,keV &     2     \\
1976-04-30  & 42898  &  150      &  2--10\,keV &     2     \\
1976-11-03  & 43085  &  90       &  2--10\,keV &      2     \\
1977-12-07  & 43484  &  600      &  2--10\,keV &      3     \\
1978-05-21  & 43649  & 160       &  2--10\,keV &      2     \\
1983-06-26  & 45511  &  35       &  2--10\,keV &      4     \\
1984-07-01  & 45882  &  52       &  2--10\,keV &      4     \\
1985-01-04  & 46069  &  41       &  2--10\,keV &      4     \\
1992-03-03  & 48684  &  >21      &  20--50\,keV &      5     \\
1992-09-07  & 48872  &  80       &  20--50\,keV &      5     \\
1993-03-18  & 49064  &  120      &  20--50\,keV &      5     \\
1993-09-20  & 49250  &  100      &  20--50\,keV &      5     \\
1994-03-21  & 49432  &  550      &  20--50\,keV &      5     \\
1994-09-21  & 49616  &  134      &  20--50\,keV &      5     \\
1995-04-01  & 49808  & 180       &  20--50\,keV &      5     \\
1995-10-02  & 49992  & 70        &  20--50\,keV &      5     \\
1996-04-07  & 50180  & 105       &  20--50\,keV &      5     \\
1996-10-18  & 50364  &    550    &  20--50\,keV &      5     \\
1997-04-12  & 50550  &   94      &  20--50\,keV &      5     \\
1997-10-15  & 50736  &   >19     &  20--50\,keV &      5     \\
2003-05-30  & 52789  &   33      &  22--40\,keV &      6     \\
2004-06-04  & 53160  &   15      &  22--40\,keV &      6     \\
2005-06-18  & 53539  &   7       &  22--40\,keV &      6     \\
2015-03-04  & 57085  &   60      &  2--10\,keV  &      7     \\
2015-09-03  & 57268  &  36       &  2--10\,keV  &      6     \\
2016-09-18  & 57649  &  4        &  2--10\,keV  &      6     \\ 
2017-03-19  & 57831  &  30       & 15--50\,keV  &      6     \\
\hline                                   
\end{tabular}
\end{center}

References: \tablefoottext{1}{\cite{priedhorsky1983}}, \tablefoottext{2}{\cite{watson1981}},\tablefoottext{3}{\cite{white1980}},\tablefoottext{4}{\cite{cook1987-a}},\tablefoottext{5}{\cite{wilson-hodge1999}},\tablefoottext{6} This work,\tablefoottext{7}{\cite{mihara2015}}.

\end{table}

\subsubsection{X-ray spectrum}

During the latest period of X-ray activity, we triggered a \textit{Swift}/XRT target-of-opportunity observation of H~1145--619, which was executed on 6 September 2015. 
The observed X-ray flux was ($8.77\pm0.11) \times 10^{-10}\,{\rm erg\,cm^{-2}\,s^{-1}}$ in the 2--10 keV band ($\sim$36~mCrab).

\begin{figure}[h]
\centering
\includegraphics[width=9cm]{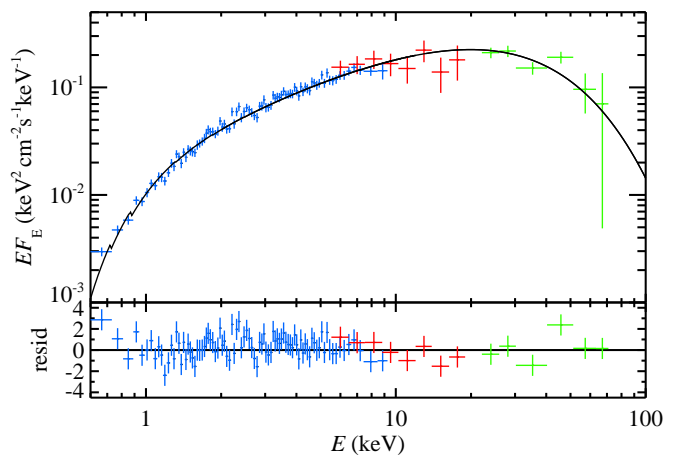}
\caption[H~1145--619 spectrum]{\textit{Swift}/XRT (blue), \textit{INTEGRAL} JEM-X (red), and ISGRI (green) spectrum of H~1145--619. The \textit{INTEGRAL} spectra were extracted from the 2003 outburst (peaking at MJD~52789) and the \textit{Swift}/XRT from the September 2015 outburst (peaking at MJD~57268). The \textit{Swift}/XRT and \textit{INTEGRAL} JEM-X spectra were normalised to the ISGRI spectrum with a normalisation constant. }
\label{fig:H1145_spectrum}
\end{figure}

As this flux level was similar to that measured in May and June 2003 by \textit{INTEGRAL}/JEM-X/ISGRI, we modelled the joint \textit{Swift}/XRT \textit{INTEGRAL}/JEM-X/ISGRI spectrum via {\sc xspec} version 12.9.0. The X-ray spectrum is shown in Fig.~\ref{fig:H1145_spectrum}. It displays a clear cut-off at high energies. Therefore, we used an absorbed cut-off power-law model to describe the spectrum; $F(E) = K E^{-(\Gamma-1)}\exp(-E/E_{\rm fold})$, $\Gamma$ is the photon index, $E_{\rm fold}$ is the e-folding energy, and $K$ is an arbitrary normalisation constant.
The fit gives $\chi^2 = 379$ for $380$ d.o.f.. We find a best-fitting folding energy of $E_{\rm fold} = 17_{-2}^{+3}$ keV and a photon index of $\Gamma = 0.86\pm0.05$.
The measured hydrogen column density towards H~1145--619, $N_{\rm H} = (0.44\pm0.04) \times 10^{22}\,{\rm cm^{-2}}$ ({\sc tbabs} model; abundances from \citealt{wilms2000}), is a factor of 3 lower than the value derived from HI maps \citep{kalberla2015}. This can be explained taking into account the distance to H~1145-619, which is estimated to be 3.1$\pm$0.5\,kpc \citep{stevens1997}. From the galaxy HI maps, we see that the main emission from the Galactic spiral arms is placed behind the source. On the other hand, the X-ray spectrum does not show any significant spectral features, such as cyclotron scattering resonance features or an iron line at 6.4~keV.

In September 2016, H~1145--619 underwent another type I outburst, which was fainter than the 2015 outbursts ($\sim$4~mCrab in the 2--10 keV band). We observed the source again with \textit{Swift}/XRT and modelled the spectrum, fixing the $E_{\rm fold}$ to 17~keV. Using the same model, we got a slightly lower column density ($N_{\rm H} = 0.22\pm0.07 \times 10^{22}\,{\rm cm^{-2}}$) and a photon index that is consistent with the value obtained in 2015 ($\Gamma = 0.80\pm0.07$).

\section{Discussion}\label{sec:dis}

We studied the long-term optical variability and X-ray activity of the Be/XB H~1145--619. We also identified several X-ray outbursts detected with \textit{INTEGRAL} and \textit{Swift}, and analysed its X-ray spectrum.

\subsection{Optical variability}\label{sub:opt_var}

As described in Sect.~\ref{subsec:opt_phot}, the optical photometric light curves of the system displayed very different patterns of variability (see Fig.~\ref{fig:opt_trends}).

From the 80s to the 90s, the optical brightness exhibited slow brightening and weakening episodes. The $V$--band brightness changed in the range 8.8--9.2 magnitudes on timescales of years. Around 1990, the source entered into a low-optical state with a smooth but continuous fading in optical brightness. In this epoch, the H$\alpha$ EW decreased significantly from EW$\simeq$-37\,\AA\, to EW$\simeq$-7\,\AA, while intense X-ray activity was taking place; H~1145--619 underwent an X-ray outburst at every periastron passage from 1991 to 1997. After that, the source went into quiescence in X-rays. This kind of variability can be explained by the evolution of the circumstellar disc of the Be star. The equivalent width of the H$\alpha$ line is considered a good indicator of the size of the disc \citep{reig2016}. The X-ray activity started in 1992 when the largest H$\alpha$ EW values were measured (EW=-37\AA), indicating that the disc was at its maximum extension. In this situation, matter from the external parts of the disc can be transferred onto the neutron star during the periastron passage, giving rise to the type I outbursts. After the successive X-ray outbursts the disc gradually shrank, as shown by the weakening of both the H$\alpha$ line and optical flux. In particular, the H$\alpha$ EW decreased from EW $\sim-37\,\AA$ (March 1994) to EW $\sim-20\,\AA$ (July 1994), after the bright outburst in March 1994 (see Tables~\ref{tab:spec_obs} and ~\ref{tab:xray_outb}). This could be due to the loss of matter transferred to the neutron star and the partial disruption of the disc produced by the X-ray emission. This behaviour has been observed for several Be/XB systems during and after the X-ray emission episodes \citep{reig2016}. It is noticeable that the decrease of the optical flux preceded the weakening of the H$\alpha$ line. The same behaviour has been observed in 2016--17; see below.

From 2001 to 2009, the optical brightness remained substantially lower than in the other periods with a mean value of $V\sim$~9.3\,mag. From 2009 to 2016, we observed a substantial increase of the optical flux ($\sim$0.5\,mag), which can be interpreted as the re-formation of the disc \citep{okazaki2001, haigh2004}. From 2015 to mid 2016, the optical magnitude remained stable around $V\sim$~8.8\,mag, which is a value close to the maximum recorded in 1990, just before the X-ray activity began \citep{stevens1997}. During 2016 the H$\alpha$ EW increased significantly to values around -25\,\AA. This increase has continued in 2017, reaching values up to -37\,\AA, which was the maximum value measured in the 90s. As expected, and coinciding with the increase of the H$\alpha$ EW, H~1145--619 entered a renewed X-ray activity phase and during the five latest periastron passages faint X-ray outbursts have been identified (see Sect. Sect.~\ref{sub:xact} and Table~\ref{tab:xray_outb}). Again, and as in the early 90s, a fading of the optical brightness preceded the occurrence of the maximum value of the H$\alpha$ EW and the onset of the X-ray activity. If the system follows the same trend, a bright X-ray outburst should happen in one of the next periastron passages.

If we compare these two events from 1990 and 2016, we can derive a long-term period P$_{long-term}\sim$~26\,yr that would be related to the formation and depletion of the circumstellar disc and to the X-ray active phases of the source. This period would also be compatible with the X-ray activity observed in the 70s.

In addition to the long-term fading and brightening of the optical emission, during its faintest state,  the source displayed superorbital modulations (see Fig.\ref{fig:opt_trends}). Although it seems this effect started coinciding with the beginning of the low-optical state around 1997, the lack of data prevented us from determining an exact date for their appearance. The good coverage of the photometric observations during the interval 2002--2009 allowed us to determine a quasi-period of P$_{superorb}\sim$~590\,d. Quasi-periodic variability with periods from a few to hundred days has been found in several isolated Be stars of early type, both in the Milky Way \citep{hubert1998,labadie-bartz2016} and in the SMC \citep{mennickent2002,mennickent2006}. Similar variations have also be seen in other Be/XBs, such as the 421\,d quasi-periodic modulation in A~0538--66, a binary with an orbital period of 16.65 d in the Large Magellanic Cloud \citep{alcock2001}. \citet{rajoelimanana2011} found superorbital quasi-periodic variability for 22 objects in a sample of 48 Be/XBs in the SMC with periods ranging from 200 to 3000 days. From the $V-$ and $I$--band light curves and the colour--magnitude diagram of this epoch, we can see that during this period in which H~1145--619 displayed superorbital variations, the source became redder when brighter (see Fig.~\ref{fig:CMD_superorb}). \citet{rajoelimanana2011} also found that for most of the Be/XBs in their sample, the colour variations were correlated with the brightness of the source, implying that these variations in the light curve are related to the behaviour of the Be circumstellar disc. Quasi-periodic variations with periods between one and ten years and the same colour-magnitude behaviour are also commonly observed in isolated Be stars \citep{jones2013}. \citet{rajoelimanana2011} proposed that superorbital photometric variations can be produced by the propagation of density waves in the Be star disc, as their characteristic timescales are very similar to those of the V/R cyclic variability. \citet{labadie-bartz2016} found a correlation between photometric and V/R variability for the Be star HD~33232. An alternative explanation could be the warping and precession of the circumstellar disc, which has been observed for other Be/XBs, such as 1A~0535+262 \citep{moritani2013}.

The third type of variability observed in the optical photometric data of H~1145--619 is characterised by optical outbursts that manifest as sudden increases and decreases in optical brightness (by $\sim$0.3\,mag). Superimposed on the superorbital variations, three optical outbursts can be observed in the ASAS light curves in the 2003--2005 period (Fig.\ref{fig:opt_trends}), preceding the three X-ray outbursts we identified with \textit{INTEGRAL} data. These outbursts are shorter than the orbital period ($\sim$80\,days), and the lags between the optical and X-ray outbursts decrease from 2003 to 2005.  Similar outbursts (in duration and intensity) were also observed in 1998–1999. These are not coincident with periastron passages and, for this reason, a relation with the interaction with the NS can be excluded. One of these outbursts was observed in two bands, the $I$ and $V$ bands. The amplitude of variation in the $I$ band is larger than in the $V$ band (see colour--magnitude diagram in Fig.\ref{fig:CMD}, right); hence, as occurred for the superorbital variations, the source is also redder when brighter during these outbursts. The same behaviour has been observed in the past for some Be stars in the Small Magellanic Cloud (SMC) (see \citet{dewit2006,rivinius2013}). These authors found that the outflowing material produces a bi-valued colour--magnitude relation (a loop in the colour--magnitude diagram), which can be attributed to optical depth effects. These loops are found to evolve clockwise for outflowing material and  counter-clockwise for accreted material. As can be seen in Fig.~\ref{fig:CMD}, we also observe a clockwise evolution in the colour--magnitude diagram, so the optical outbursts we observe in this period are probably due to ejection of material from the Be star. Mass ejections from the stellar photosphere are a common feature in Be stars in general \citep{rivinius2013} and has also been observed in some Be/XB systems \citep{yan2012, li2014}.

\subsection{V/R variability and discovery of a retrograde one-armed density wave}\label{sub:disc_denswave}

Spectroscopically, the presence of a density perturbation in the disc is revealed by changes in the relative intensity for the split profile of certain emission lines, namely V/R variability. The effect is most prominent in the H$\alpha$ line. In many of the spectroscopic observations presented in this work, V/R oscillations have been found.

{In the spectra from the 90s, asymmetric double-peaked profiles were observed, but the temporal coverage is not sufficient for a deeper analysis.

In the period from 2002 to 2009, when the 590\,d-superorbital photometric variations were observed, only two spectra were obtained (15 March 2005 and 8 May 2006). Both spectra display very asymmetric lines with different V/R values (see Table~\ref{tab:spec_obs} and Fig.~\ref{fig:spec_old}). The observed asymmetry and the change in V/R would be in agreement with the scenario we proposed in Sect.~\ref{sub:opt_var} in which the superorbital variations would be produced by the presence of a density wave. Additional data would be needed to support this hypothesis and to rule out another explanations, such as the warping of the circumstellar disc.

In the March 2010 spectrum, the H$\alpha$ and He{\sc i} lines displayed almost symmetric double-peaked profiles, which suggest that the mechanism producing the superorbital variations could have disappeared.

During the 2015 observations, the lines again showed double-peaked line profiles with V$\approx$ R, and the EW and RV values did not vary substantially, indicating that the disc remained very stable over this period (see Figs.~\ref{fig:spec_res} and \ref{fig:spec_RVs}).

However, the V/R variability returned in 2016 and it is still present in July 2017. As shown in Sect.~\ref{sub:opt_spec}, we identified V/R oscillations in the 2016--2017 optical spectra. Apart from this, we also observed quasi-periodic variability in the radial velocities (and in $\Delta$V) of the peaks, which could be interpreted as evidence of the presence of a one-armed density wave. \citet{telting1994} presented observations of $\beta^1$~Mon in which they identified the presence of a prograde one-armed density structure in the circumstellar disc material. Observations of prograde one-armed oscillations have been also described by \citet{reig1997} and \citet{mennickent1997} in the past. For H~1145--619, we observed that the one-armed oscillation moves in the opposite direction as the material on the near-Keplerian circumstellar disc (see Fig.~\ref{fig:wden}). As in \citet{telting1994}, we also identified four states in the V/R cycle, but with a different evolution as follows:

\begin{itemize}
 \item (I) Blue dominated profiles (V/R$>$1): when the density enhancement is on the side of the disc that is moving towards the observer. We saw this behaviour in our spectra from February to April 2016, and from April to May 2017.
 \item (II) Symmetric profiles with V/R$\approx$1 and no shell absorption: when the density enhancement is behind the star. We observed a single-peaked profile, which could be explained as a particular case of this situation in which the inclination is small enough to see most of the enhanced emission, which would be RV$\approx$0~km/s. We observed this feature in May 2016 and in July 2017.
 \item (III) Red dominated profiles (V/R$<$1): when the density enhancement is on the side of the disc moving away from the observer. We observed this effect in June--July 2016.
 \item (IV) Shell phase, symmetric double-peaked profiles (V/R$\sim$1) with central shell absorption: when the density enhancement is between the star and the observer. We observed signatures of this situation in January--February 2017. In our case, the shell feature is more pronounced for the He{\sc i} line; this makes sense for a moderate inclination, which also needed to explain the single-peaked profiles on state II. To observe shell absorption, the inclination should be large enough, but as explained above, it cannot be 90\degr. In Sect.~\ref{sec:sptype}, we estimated the inclination to be close to $i \approx $70\degr, which is compatible with this scenario.
 \end{itemize}
According to \citet{telting1994}, the evolution of the line profiles described above (I--II--III--IV) corresponds to the density wave that is moving in the opposite direction to the material in the circumstellar disc (see Fig.~\ref{fig:wden}). A prograde wave would produce the sequence IV--III--II--I with red dominated profiles after the shell phase.

Retrograde one-armed oscillations have been predicted in the past \citep{okazaki1985,okazaki1997}, and according to recent studies, the simplest analysis points to the fact that discrete one-armed oscillation modes are retrograde and not prograde \citep{kato2016}. However, as far as we know, no observational evidence of their existence had been previously observed. From the occurrence of the two similar single-peaked profile states (in May 2016 and July 2017), we can infer a tentative period of the rotation of the density wave around the star of P$_{dens-wave}\sim$~420\,d.This period is rather short when compared with periods observed in isolated Be stars, which typically range between five to ten years \citep{okazaki1997}. However, it has been shown that the V/R quasi periods of Be star discs in Be/XB systems are significantly shorter \citep{clark1998, reig2010} owing to the smaller size of the discs in these systems \citep{reig2016}.

In Sect.~\ref{ssub:BD}, we presented the evolution of the Balmer decrement D$_{34}$ during 2015 and 2016. This points to an increase of the density in 2016, which we think is related with the development of the density wave.

\subsection{Identification of new X-ray outbursts and analysis of the X-ray spectra}

As shown in Sect.~\ref{sub:xact}, the analysis of \textit{INTEGRAL} archival data identified three X-ray outbursts that took place in 2003, 2004, and 2005. The three X-ray outbursts occurred a few days before the corresponding periastron passages. This fact is in agreement with the behaviour of the outbursts detected by \citet{wilson-hodge1999} using \textit{CGRO}/BATSE data, which took place within a narrow range of orbital phase (-0.1--0.1). Two of these outbursts have been mis-identified as {\sl low luminosity} outbursts from the close system 1E~1145.1--6141, which is at $\sim$~17~arcmin from H~1145--619 \citep{ferrigno2008}, while the 2003 outburst was associated with H~1145--619 in \citet{filippova2005}. We also identified two faint X-ray outbursts during the periastron passages in September 2016 and in March 2017.

For the analysis of the X-ray spectrum, we modelled our X-ray spectra with an absorbed cut-off power-law model, and we obtained $E_{\rm fold} = 17$~keV and a photon index of $\Gamma = 0.86$. \citet{white1980} analysed the X-ray spectra of 2 August 1978 by HEAO--1 (MED and HED) and obtained $\Gamma \sim 1.5$ and no signs of a high-energy cut-off below 60~keV or an iron line at 6.7~keV. Such a featureless power-law spectrum is typical from the emission originated by matter falling onto a magnetised neutron star. On the other hand, neither signatures of emission and reflection by an accretion disc nor cyclotron lines can be identified on the spectrum. On the other hand, \textit{Ariel~V} obtained a spectrum in December 1978 that is well fitted with a power law of index $\Gamma = 1.0$ and a high-energy cut-off at 8.5~keV \citep{white1978}. This result is similar to the values obtained in \citet{cook1987-a} with \textit{EXOSAT}/ME data of the July 1984 and January 1985 outbursts, who obtained several spectra with $\Gamma \sim 1.0$ and high-energy cut-off at $\sim$6~keV. We should consider the proximity of the pulsar 1E~1145.1--616 when comparing these results with ours. 1E~1145.1--616 has shown high column density ($N_{\rm H} \sim 10^{23} $cm$^{-2}$;\citealt{white1980, ferrigno2008}), so contamination from this source could result in larger values of the photon index. When the \textit{HEAO--1} observations were performed, 1E~1145.1--616 was off, while in the \textit{Ariel~V} observations it was detected. In the \textit{EXOSAT} observations, there is little evidence of a significant contribution to the measured flux from 1E~1145.1--616. However, contamination from 1E~1145.1--616 should not be significant in our \textit{Swift}/XRT and \textit{INTEGRAL}/JEM--X observations, and we have tried to avoid it in the analysis of the \textit{INTEGRAL}/IBIS data.

\section{Conclusions}\label{sec:con}

We have performed a multiwavelength analysis of the Be/XB H~1145--619 covering more than 40 years of data, from 1973 to 2017, with the goal of studying the correlations between the optical and X-ray variability.

A new spectral classification of B0.2\,III for the optical companion of the system is proposed. A value of v~$\sin i$ = 300~km/s is also determined. Considering a critical fraction of $w$=0.8, we derive an inclination of the circumstellar disc $i\simeq$\,70\degr.

From the optical observations, we found that the circumstellar disc has suffered enhancement and weakening episodes in the 80s and 90s and later from 2010 to 2017. In the period 2000--2009, the optical brightness was fainter than in the previous decade by $\sim$~0.5\,mag and the source exhibited superorbital variations with a period of P$_{superorb}\sim$~590\,d, which could be explained by the propagation of density waves in the disc or alternatively by warping of the disc. Moreover, optical outbursts were observed in 1998, 1999, 2003, 2004, and 2005, which we attributed to mass ejections from the Be star.

From the optical spectra, we discovered hints of the presence of density waves during the 90s and 00s. In 2010, no asymmetries were found. From the spectra acquired from 2015 to 2017, we have identified the absence of asymmetries in 2015, and the appearance in 2016 of a retrograde one-armed oscillation in the circumstellar disc. In the latest spectra from July 2017, this density wave is still present and appears brighter than in 2016, reaching values of the EW closer to the brightest density wave observed in the 90s. We estimated a tentative period for the motion of the density wave during 2016--2017 of P$_{dens-wave}\sim$~420\,d. Retrograde one-armed density waves are predicted by the theoretical models, but this is the first time that spectroscopic evidence of these waves in a Be X-ray binary system are observed.

We identified three X-ray outbursts in 2003, 2004, and 2005 with \textit{INTEGRAL}/IBIS and JEM--X data. Since March 2015, the source underwent four normal X-ray outbursts at all but one periastron passage. We also modelled the X-ray spectrum using \textit{INTEGRAL} and \textit{Swift}/XRT data from 2003 and 2015, respectively, and we obtained the best fit with an absorbed cut-off power-law model with folding energy of $E_{\rm fold} = 17_{-2}^{+3}$ keV and a photon index of $\Gamma = 0.86\pm0.05$.

As already noted in Sect.~\ref{sec:dis}, given the similarities with the active period of the 1990s and taking the presence of the density wave and the observed increase of the electron density in the disc into account, we expect the X-ray activity to continue with a bright X-ray outburst probably taking place in the coming years. For this reason, monitoring the source properly is crucial. We have approved \textit{INTEGRAL} and \textit{Swift} ToO proposals in case this system undergoes a new outburst in 2017 and/or 2018 and we will continue the optical monitoring to continue studying the evolution of the disc.

\begin{acknowledgements}

The authors would like to thank the anonymous referee for his/her very helpful comments and suggestions on the original manuscript. This work has been supported by Spanish MINECO grant ESP2015-65712-C5-1-R. This research has made use of data from the \textit{INTEGRAL}/OMC Archive at CAB (INTA--CSIC), pre-processed by ISDC. The work of J.F. is supported by the Spanish Ministerio de Econom\'\i a y Competitividad, and FEDER, under contract AYA2013-48623-C2-2-P, and by the Generalitat Valenciana project of excellence PROMETEOII/2014/060. JJEK acknowledges support from the Academy of Finland grants 268740 and 295114, the V\"{a}is\"{a}l\"{a} Foundation, and the ESA research fellowship programme. LJT is supported by the University of Cape Town Research Committee. We thank the V\"{a}is\"{a}l\"{a} Foundation and the Faculty of the European Space Astronomy Centre for their support to obtain the \textit{SMARTS}/CHIRON observations used in this research. Some of the observations reported in this paper were obtained with the Southern African Large Telescope (SALT) under proposal 2015-1-SCI-032 and 2016-1-DDT-001. 

\end{acknowledgements}

\bibliographystyle{bibtex/aa.bst} 
\bibliography{allcites} 

\end{document}